\documentclass[journal]{IEEEtran}
\IEEEoverridecommandlockouts
\usepackage{cite}
\usepackage{amsmath,amsfonts,amssymb,amsthm}
\usepackage{algorithmic}
\usepackage{graphicx}
\usepackage{textcomp}
\usepackage{xcolor}
\usepackage[utf8]{inputenc}
\usepackage{commath}
\usepackage{scalerel}
\usepackage{tikz}
\usepackage[]{changes}
\usepackage{caption}
\usepackage{subcaption}
\usepackage{layouts}
\usepackage[draft]{hyperref}
\usepackage{todonotes}
\usepackage{balance}
\usepackage{textcomp} % nice greek alphabet
\usepackage{pifont}   % Dingbats
\usepackage{booktabs}
\usepackage{amssymb,amsthm,amsfonts}
\usepackage{amsmath}
\usepackage{subcaption}
\usepackage{tcolorbox}
\usepackage{environ}
\usepackage{rotating}

\tikzset{font={\fontsize{10pt}{12}\selectfont}}
\usepackage{pgfplots}
\DeclareUnicodeCharacter{2212}{−}
\usepgfplotslibrary{groupplots,dateplot,polar}
\usetikzlibrary{patterns,shapes.arrows}

\pgfplotsset{compat=newest}

\definecolor{color2}{RGB}{85,168,104}
\definecolor{color1}{RGB}{221,132,82}
\definecolor{color0}{RGB}{76,114,176}

\def\mystrut{\vphantom{hg}}

\pgfplotsset{
    legend image with text/.style={
        legend image code/.code={%
            \node[anchor=center] at (0.3cm,0cm) {#1};
        }
    },
}

\usepackage{environ}
\makeatletter
\newsavebox{\measure@tikzpicture}
\NewEnviron{scaletikzpicturetowidth}[1]{%
  \def\tikz@width{#1}%
  \def\tikzscale{1}\begin{lrbox}{\measure@tikzpicture}%
  \BODY
  \end{lrbox}%
  \pgfmathparse{#1/\wd\measure@tikzpicture}%
  \edef\tikzscale{\pgfmathresult}%
  \BODY
}
\makeatother

\def\BibTeX{{\rm B\kern-.05em{\sc i\kern-.025em b}\kern-.08em
    T\kern-.1667em\lower.7ex\hbox{E}\kern-.125emX}}

\begin{document}

\title{3D Non-Stationary Channel Measurement and Analysis for MaMIMO-UAV Communications
%Measurement-based UAV Air-to-Ground Channel %3D
%Stationarity Characterization
%\thanks{}
}

% \and
% \and

\author{
Achiel Colpaert,~\IEEEmembership{Member,~IEEE}, Zhuangzhuang~Cui,~\IEEEmembership{Member,~IEEE}, \\Evgenii~Vinogradov,~\IEEEmembership{Member,~IEEE}, and Sofie Pollin,~\IEEEmembership{Senior Member,~IEEE}
%\thanks{This research is supported by the Research Foundation Flanders (FWO) project no. G098020N.}
\thanks{
Copyright (c) 2015 IEEE. Personal use of this material is permitted. However, permission to use this material for any other purposes must be obtained from the IEEE by sending a request to pubs-permissions@ieee.org.\\
This research is supported by the Research Foundation Flanders (FWO) project no. G098020N and the iSEE-6G project under the EU’s Horizon Europe Research and Innovation programme with Grant Agreement No. 101139291. \textit{(Corresponding authors: Achiel Colpaert, Zhuangzhuang Cui)}}%\zh{do}}
\thanks{A. Colpaert is with IMEC, Kapeldreef 75, 3001 Leuven, Belgium, and also with WaveCoRE of the Department of Electrical Engineering (ESAT), KU Leuven, Leuven, Belgium. E-mail: achiel.colpaert@imec.be.}
\thanks{Z. Cui, S. Pollin are with WaveCoRE of the Department of Electrical Engineering (ESAT), KU Leuven, Leuven, Belgium. E-mail: \{zhuangzhuang.cui, sofie.pollin\}@kuleuven.be}
\thanks{E. Vinogradov is with ESAT, KU Leuven, Leuven, Belgium, also with Autonomous Robotics Research Center, Technology Innovation Institute (TII), Abu Dhabi, UAE. E-mail: evgenii.vinogradov@tii.ae.}
}

\maketitle

\begin{abstract}
Unmanned aerial vehicles (UAVs) have gained popularity in the communications research community because of their versatility in placement and potential to extend the functions of communication networks. However,  there remains a gap in existing works regarding measurement-verified stationarity analysis of the air-to-ground (A2G) Massive Multi-Input Multi-Output (MaMIMO) channel which plays an important role in realistic UAV deployment. In this paper, we first design a UAV MaMIMO communication platform for channel acquisition. We then use the testbed to measure uplink Channel State Information (CSI) between a rotary-wing drone and a 64-element MaMIMO base station (BS). 
For characterization, we focus on multidimensional channel stationarity which is a fundamental metric in communication systems. Afterward, we present measurement results and analyze the channel statistics based on power delay profiles (PDPs) considering space, time, and frequency domains. We propose the stationary angle (SA) as a supplementary metric of stationary distance (SD) in the time domain. We analyze the coherence bandwidth and RMS delay spread for frequency stationarity. Finally, spatial correlations between elements are analyzed to indicate the spatial stationarity of the array. The space-time-frequency channel stationary characterization will benefit the physical layer design of MaMIMO-UAV communications. 
\end{abstract}


\section{Introduction}
\label{sec:introduction}

Recently, Morgan Stanley released a forecast stating that by 2050 the total market of Urban Air Mobility (UAM), including drone delivery, air taxi, and patrolling drones, \textit{etc.}, will reach up to 11\% of the projected global Gross Domestic Product (GDP) \cite{MS}. It also claims that large-scale deployments of drone-based urban delivery are expected to be in practice by 2030. Moreover, a short-term prediction states that the drone services market will be worth a total of USD 63.6 billion by 2025 \cite{business2023drone}. %, and the UAV fleet, both recreational and commercial, is projected to reach 2--3 million by 2023 \cite{faa2019prediction}. 
Apart from UAM applications, Unmanned Aerial Vehicles, and Systems (UAVs, UASs) are expected to play an essential role in the sixth-generation (6G) networks \cite{Tut18, azari2021evolution, UAV_6G}.% where aerial base stations (BSs) can provide ubiquitous connectivity to ground users \cite{Tut18, azari2021evolution, UAV_6G}. 

Ensuring the expected levels of safety, security, operation transparency, and airspace usage efficiency of such a large fleet, requires seeking UAS Traffic Management (UTM) solutions %\cite{Dr_Tech}. 
The UTM framework formulated by the International Civil Aviation Organization (ICAO) \cite{ICAO} includes cellular networks as a critical enabler of large-scale drone deployments. Although several competing UTM implementations have been proposed \cite{Bauranov}, all of them rely on a reliable command and control (C2) link. It has been acknowledged that the cellular network is an ideal candidate as it provides ubiquitous coverage and removes the need for the UTM provider to deploy expensive dedicated wireless infrastructure. Therefore, the 3rd Generation Partnership Project (3GPP) has taken the first step to embed aerial vehicles into cellular networks \cite{3gpp2018enhanced}. 
%\zc{the first two paragraphs can be merged.}

%\zc{StoA for cellular-connected UAV communications: }

%\zc{
For cellular-connected UAV communications, most of the existing works focus on performance analysis, trajectory design, and networking design \cite{zengccuc}. In \cite{mamimouavmag}, the authors study the Massive Multi-Input Multi-Output (MaMIMO)-supported UAV communications, where the fine-grained 3D beamforming is used to mitigate the interference between UAVs and terrestrial users.
In our previous works, we investigated aerial coverage analysis using LTE BS down-tilted pattern in a realistic 3D city model \cite{colpaert18}, and 3GPP antenna and channel models \cite{cui22globecom}. Moreover, UAV mobility causes significant handover challenges analyzed and tackled in \cite{colpaert2020beamforming,bernabe2023}. 
As one of the main methodologies in channel modeling, geometry-based stochastic modeling (GBSM) has been widely used to model the UAV channels in a multi-antenna system \cite{jiang18}. The scatterers are assumed to be distributed in a specific geometry, such as cylinder, sphere, and ellipsoid \cite{jiang21}. The method results in a better generality, however, more measurement verification is needed. In terms of channel measurement and characterization, authors in \cite{amorim2017radio, al2017modeling} measured LTE-connected UAV channels, where large-scale fading, i.e., path loss and shadowing models are the main focuses. Moreover, most existing measurement campaigns are based on a single antenna \cite{9107106,8770066,10001404,8766809,9448737, 8741964}, and few of them used a very limited number of antennas to study the MIMO-UAV channels such as eight antennas in \cite{7079507}. 
Furthermore, due to the high mobility and different environment of UAVs, we expect a significant difference in terms of channel stationarity. The UAV MaMIMO channel will have a larger temporal stationarity since the channel is LOS dominated. State-of-the-art has already indicated that MaMIMO channels are more deterministic, thus it becomes interesting to perform a location-based evaluation of the stationarity regions of such channel. More specifically it has been shown the angular properties of the MaMIMO channel are a powerful tool to characterize the channel \cite{ali18millimeter}.
The non-stationarity plays an important role in practical communication systems, such as reconfigurable intelligent surface (RIS)-assisted communication, where the spatiotemporal channel dynamics and near-field effects impact the resultant gain generated from the RIS \cite{jiangris1,jiangris2}. However, the study of non-stationarity of MaMIMO-connected UAV communication is still in its infancy, in which existing works concentrate on theoretical analysis of its feasibility \cite{Geraci18}. For MaMIMO to UAV channels, authors in \cite{bai2022non} adopted the GBSM to characterize the space-time-frequency non-stationarities. In addition, a beam-domain channel model is proposed for analyzing the multidimensional channel correlation \cite{chang2023novel}. These theoretical works provide comprehensive channel analysis, however, based on various assumptions. Moreover, the proposed models are too complicated to be applied in reality. Therefore, realistic measurements using a MaMIMO testbed for UAVs are desperately needed, besides, a detailed characterization of channel correlation and stationary characteristics will be more beneficial to the design and optimization of the physical layer. To our best knowledge, it is the first work to employ a realistic MaMIMO testbed for measuring UAV air-to-ground (A2G) channels.

However, to conduct the UAV-based A2G channel measurement using the MaMIMO testbed, many open issues remain to be addressed \cite{Tut18, UAV_6G}. %which are as follows. 
%\subsection{State of the Art: overview of open problems}
The first issue is the \textit{measurement system design}. In existing work, most measurement systems utilize a single antenna at both transmitter (Tx) and receiver (Rx) \cite{cui22}. With this setup, large-scale or narrow-band channel characteristics are generally analyzed, such as path loss, shadow fading, and small-scale fading statistics. However, spatial channel characteristics are not able to be investigated. Nonetheless, current work with multiple-antenna setup mostly focuses on the distribution of multipath components (MPCs), where a group of angular spectrum, e.g., angle of arrival (AoA) and angle of departure (AoD), are popular metrics. The temporal channel correlation is generally ignored due to the quasi-static MaMIMO measurement system. Thus, to study multidimensional channel stationarity, a measurement system should have massive antennas, use enough bandwidth, and support mobile aerial users. %\{maybe introduce TDD and OFDM waveform here (why we use this system compared to existing ones), then we can say OTA difficulty\}. 
Then, Over-the-Air (OTA) synchronization plays an important role in accurately obtaining the channel state information (CSI). For measuring A2G channels where UAV flies in the sky, it is tricky to realize a perfect synchronization due to the high mobility. Moreover, in the time-division duplexing (TDD) system, time synchronization becomes more challenging. Thus, our first focus aims at the measurement system design to facilitate CSI acquisition.

Afterward, more effort is needed in characterizing \textit{multidimensional channel stationarity}. For a time-varying scattering environment due to the mobility of drones, it is evident that MPCs change with time. Therefore, authors in \cite{he2015characterization}
%\{cite ruisi's paper\}
analyzed the stationary distance as a critical metric to measure the channel stationarity in the time domain for vehicle-to-vehicle (V2V) channels. The stationary characteristics are an important reference to the system design and performance evaluation of corresponding networks. Existing works in \cite{kaili18, junw21} %\{cite some papers using GBSM that analyzes 3D stationarity\} 
incorporated the channel stationarity into the general GBSM model, where the scatterers are modeled in a stochastic way and the stationarity is assumed to be independent among time, frequency, and space domains. Most works study the non-stationarity of UAV-MIMO channels based on the GBSM \cite{8594724,8340788,9552011}. Thus, there is a gap in characterizing wireless channel stationarity based on measurements for MaMIMO-UAV communications, which motivates our second focus, i.e., multidimensional channel stationarity characterization.

In this paper, we first develop a MaMIMO testbed for aerial users, which supports drone-based A2G channel measurement. We then focus on the channel stationarity analysis in terms of time, frequency, and space domain. %which aims to provide overall stationary characteristics of A2G channels. 
To our best knowledge, it is the first study that focuses on measurement-based stationary analysis for MaMIMO-connected UAV communications. 
Our main contributions are summarized as follows.  
\begin{itemize}
    \item \textbf{Measurement system design:} we develop a measurement testbed that supports multi-antenna and wideband measurement campaigns, with well-design time-frequency OTA synchronization capable of temporal stationary measurement of A2G channels using a drone. Thus, a space-time-frequency measurement platform consists of MaMIMO BS and a drone.
    \item \textbf{Channel data collection:} we conduct numerous measurements to obtain a large volume of data set considering different heights of UAV at 8~m, 11~m, and 24~m, diverse distances between UAV and BS, massive antennas (64 elements), and more usable bandwidth (18~MHz), which enables us to analyze the space-time-frequency characteristics of the A2G propagation channels.
    \item \textbf{Channel stationarity analysis:} we first show measurement results with the received power and power delay profile (PDP). For channel stationary characterization, we focus on stationary distance/angle, coherence bandwidth/root-mean-square (RMS) delay spread, and spatial correlation among elements from time, frequency, and space aspects, respectively.
\end{itemize}

The remainder of the paper is structured as follows. Section II introduces the backgrounds and metrics of channel stationarity in terms of space-time-frequency domains. Section III provides a detailed description of the measurement setup including system, scenario, and campaign. In Section IV, the measurement results are presented and the channel stationarities are analyzed. Finally, Section V concludes the paper.

%\zh{I feel that sota overview is a bit narrow. for instance, there are a lot of measurement-based works by Zach \cite{9107106,8770066}, Yang \cite{10001404}, or other people \cite{8766809,9956936, 9448737, 8741964} that probably deserve to be mentioned/described. it is necessary to explain how this paper goes beyond. also there're a few worthy theoretical papers dealing with non-stationarity \cite{8594724,8340788} or even \cite{9687128, 9552011} stating that this is an important issue}

%\zc{Maybe you could include a section to introduce the theory of channel stationarity before the measurement section.}
\section{Channel Stationarity Characterization}
\label{sec:channel}
% \zh{from this section, I did not understand what stationarity is and why it is important to consider/model. maybe you can adapt the following text?}
% UAVs operate in a 3D space with varying height, speed and environment, which result in a time- and frequency-varying channel. Combined with a MaMIMO system which introduces spatial variation of the channel between each antenna element, results in a channel that can be characterized in a 3D stationary analysis.

\par When UAV meets MaMIMO, the corresponding channel behaviors become very complicated. Firstly, UAVs fly in 3D space with varying heights and velocities, which results in time-varying multipath propagation that is height- and speed-dependent. Moreover, with a large number of antenna elements in the MaMIMO, the channel distribution becomes element-dependent. MaMIMO-UAV channels are highly non-stationary in space, time, and frequency domains. 
When a A2G MaMIMO channel is compared to a standard outdoor SISO channel main differences exists. First, the A2G MaMIMO channel is largely LOS dominated and we can consider a high probability of LOS. There are no nearby scatterers to the UAV and most reflections occur close to the ground station. Thus the A2G will have a larger temporal stationarity as there will be fewer multi-path components. Secondly, as the UAV channel will be more mobile than traditional channel thus

To better understand the non-stationary channel behavior, researchers assume that the radio channel is Wide Sense Stationary, has Uncorrelated Scattering and is Homogeneous (WSS-US-H). When the channel fulfills the listed assumptions, it is regarded to be WSS-US-H, and its statistics do not change over either time, frequency, or angle. In this case, we can create a statistical channel model that accurately reproduces the channel behavior \cite{molish2011wireless}. These assumptions cannot be applied to all existing channels and environments. However, the channel statistics do not change instantaneously. Hence, it is possible to identify the bounded time, frequency range, and distance (or an angular range) with constant statistics (i.e., a multi-dimensional region where the WSS-US-H assumption is fulfilled locally).

From propagation theory to radio engineering, in practical communication systems, channel quasi-stationarity refers to the radio channel in its current state being similar enough when compared to the channel in the neighbouring time slots, frequencies, as well as antenna elements \cite{he2015characterization}. The ranges within which this similarity of the channel holds correspond to the temporal, frequency, and spatial quasi-stationary regions. In the remainder of this work quasi-stationarity will be referred to as stationarity. 
The quantification of these slight changes between different channels can be performed by channel correlation. The characterization can be assessed by means of correlations as in \cite{he2015characterization}, however in this measurement campaign the $M \times 1$ channel frequency response vector $\textbf{H}(t,\Delta f)$ at $M$th Rx antennas at time $t$ and at frequency bin $\Delta f$.
The channel impulse response matrices can be obtained from the channel frequency response by performing an inverse fast Fourier transform resulting in the impulse response vector $\textbf{h}(t,\tau)$, where $\tau$ is the delay. 
From the impulse response, we obtain the instantaneous PDP as follows,
\begin{equation}
\label{eq:pdp}
    \textbf{P}(t,\tau) = |\textbf{h}(t,\tau)|^2,
\end{equation}
where the time $t$ can be replaced by discrete sample $n$ with a continuous-time sampling. 

%\zc{I think the details of OFDM parameters can be given in the Measurement Section.}\zh{indeed, a table with the measurements setup (frequencies, bandwidths, subcarrier spacings, sampling rate, etc) would be helpful}
%\zc{a table needed for measurement setup maybe.}

\subsection{Temporal Stationarity}
Temporal stationarity suggests the duration of channel similarity in time, which is also quantified by the stationary distance when the speed of the user is known.
The stationary distance can be obtained by the correlation matrix distance (CMD), which can be used to evaluate whether the spatial structure of a channel changes over time. 
The CMD between two correlation matrices $\textbf{R}(t_i)$ and $\textbf{R}(t_j)$ at two different times $t_i$ and $t_j$ is defined as follows:
\begin{equation}
\label{eq:cmd}
    d_{corr}(t_i,t_j) = 1 - \frac{tr\{\textbf{R}_a(t_i) \textbf{R}_a(t_j)\}}{||\textbf{R}_a(t_i)||_f  ||\textbf{R}_a(t_j)||_f},
\end{equation}
where $i$ and $j$ are time index indications, $\textbf{R}_a(t_i)$ is the averaged receive antenna correlation matrix at time $i$, and $||\cdot||_f$ is the Frobenius norm. The averaged antenna correlation matrix $\textbf{R}_a(t_i)$ is calculated as follows:
\begin{equation}
    \textbf{R}_a(t_i) = \frac{1}{BW} \sum_{\Delta f=-B/2}^{B/2} \sum_{k=i}^{i+W-1} \textbf{h}^f(t_k,\Delta f)\textbf{h}^f(t_k,\Delta f)^H,
\label{eq:Rant}
\end{equation}
where $B$ is the bandwidth within which frequency quasi-stationarity can be assumed, $W$ is the averaging window length in time, and $(\cdot)^H$ is the Hermitian or conjugate transpose of a matrix or vector. The size of $W$ needs to be chosen such that during this time-interval quasi-stationarity can be assumed. 
Then, the stationary distance is defined by the region over which the CMD stays below a certain threshold, $c_{th}$ \cite{he2015characterization}:
\begin{equation}
\label{eq:sd}
    d_{SD} = v(t_{max} - t_{min}),
\end{equation}
where $v$ is the average speed of the transmitter, and $t_{max}$ and $t_{min}$ are the bounds of the quasi-stationarity interval, given by
\begin{equation}
\label{eq:sd_int}
\begin{split}
    & t_{min} = \operatorname*{arg\,max}_{0 \leq j \leq i-1} d_{corr}(t_i,t_j) \geq c_{th}, \\
    & t_{max} = \operatorname*{arg\,min}_{i+1 \leq j \leq T - W} d_{corr}(t_i,t_j) \geq c_{th}, \\
\end{split}
\end{equation}
where $T$ is the total measurement duration. Considering a supplementary metric, the distance to the base station can be replaced by the angle to the base station, knowing the straight-line trajectory. This enables us to obtain stationary angles from equations \ref{eq:cmd}-\ref{eq:sd_int}. Section \ref{sec:sa} details more discussion and comparisons between stationary distance and stationary angle.  
%We introduce another metric more suited for MaMIMO channel characterization, namely the stationary angle (SA) which is further detailed in section \ref{sec:sa}. 

\subsection{Frequency Stationarity}
To characterize the frequency stationarity, both the RMS delay spread and coherence bandwidth can be used.  The RMS delay spread is the normalized second-order central moment of the PDP, and the coherence bandwidth is obtained by seeking the continuous similarity of channel subcarriers. Firstly, the RMS delay spread can be calculated by \cite{molish2011wireless}:

\begin{equation}
\label{eq:rms}
    S_{\tau}(t_i) = \sqrt{\frac{\sum^{N_{\tau}}_{\tau=0}P_h(t_i,\tau)\tau^2}{P_m(t_i)}-T_{m}(t_i)^{2}},
\end{equation}
where $P_m(t_i)$ is defined as:
\begin{equation}
    P_m(t_i)=\sum_{\tau = 0}^{N_{\tau}}P_h(t_i,\tau),
\end{equation}
and $T_m(t_i)$ is defined as:
\begin{equation}
    T_m(t_i) = \frac{\sum^{N_{\tau}}_{\tau=0}P_h(t_i, \tau)\tau }{P_m(t_i)},
\end{equation}
where $P_h$ is the averaged PDP over an averaging window $W$ and over all $M$ receiver antennas, defined as follows:
\begin{equation}
    P_h(t_i, \tau) = \frac{1}{MW} \sum_{k = i}^{i+W-1} ||\textbf{P}(t_k,\tau)||_1.
\end{equation}
The coherence bandwidth is obtained from the time frequency correlation function $R_f$ as follows \cite{molish2011wireless}:
\begin{equation}
\begin{split}
    B_{coh}(t_i)  =  & \frac{1}{2} \left[ \operatorname*{arg\,max}_{\Delta f > 0} \left( \frac{|\textbf{R}_f(t_i,\Delta f)|}{\textbf{R}_f(t_i,0)}=\frac{1}{e} \right) \right. \\
    & \left. - \operatorname*{arg\,min}_{\Delta f < 0} \left( \frac{|\textbf{R}_f(t_i,\Delta f)|}{\textbf{R}_f(t_i,0)}=\frac{1}{e} \right) \right]
\end{split}
\end{equation}
where $\textbf{R}_f$ is obtained by correlating the channel frequency response $\textbf{h}^f(t_i,\Delta f)$ with respect to the frequency and the threshold $1/e$ as used by He \textit{et al.}\cite{he2015characterization}.

%\ac{Another figure maybe illustrates the relationship between coherent bandwidth and RMS delay spread, with different thresholds.
%\begin{equation}
%    B_{\rm coh}=\frac{1}{\alpha S_\tau},
%\end{equation}
%where $\alpha$ represents the coefficient and $S_\tau$ is the RMS delay spread.
%\begin{equation}
%    B_{\rm coh}=\frac{1}{T}\frac{1}{N}\sum_{t=0}^T\sum_{n=1}^{N} \arg_{\Delta f}
%\end{equation}
%}

\subsection{Spatial Stationarity}
The spatial stationarity can be indicated by the channel similarity among antenna elements at the BS end. This similarity can be observed by directly evaluating the antenna correlation matrix calculated in equation (\ref{eq:Rant}) which is of shape M-by-M. Each row indicates the correlation value towards all antenna elements.

%\zc{Here I think it is just your $\textbf{R}_H$, to be checked.}
\section{Measurement Setup}
\label{sec:measurements}
This section describes the overall measurement setup. We first introduce the measurement system, both hardware and software aspects. Then, we provide a detailed description of measurement scenarios. 
\subsection{Overview of A2G Channel Measurement System}
%To capture the Air-to-Ground MIMO channel a novel measurement system was built based on the Massive MIMO testbed at KU Leuven. Full details can be found in the work of \ac{reference cming} 
In state-of-the-art four main techniques exist to perform channel sounding: RF pulse systems \cite{guanmea}, vector network analyzers, spread-spectrum real-time or sliding correlators, and orthogonal frequency-division multiplexing (OFDM) based systems \cite{MacCartney17Flexible}. Most of the MIMO channel-sounding setups have only a limited number of RF chains and rely on an RF switch to measure a large virtual MIMO array. However, while this method is sufficient to measure large-scale channel parameters, it lacks the precision required to capture channel variations between antenna elements and lacks detailed phase information between antenna elements.

\begin{figure}
    \centering
    \begin{subfigure}{\linewidth}
        \centering
        \begin{tikzpicture}
            \node[anchor=south west,inner sep=0] at (0,0) {\includegraphics[width=\textwidth]{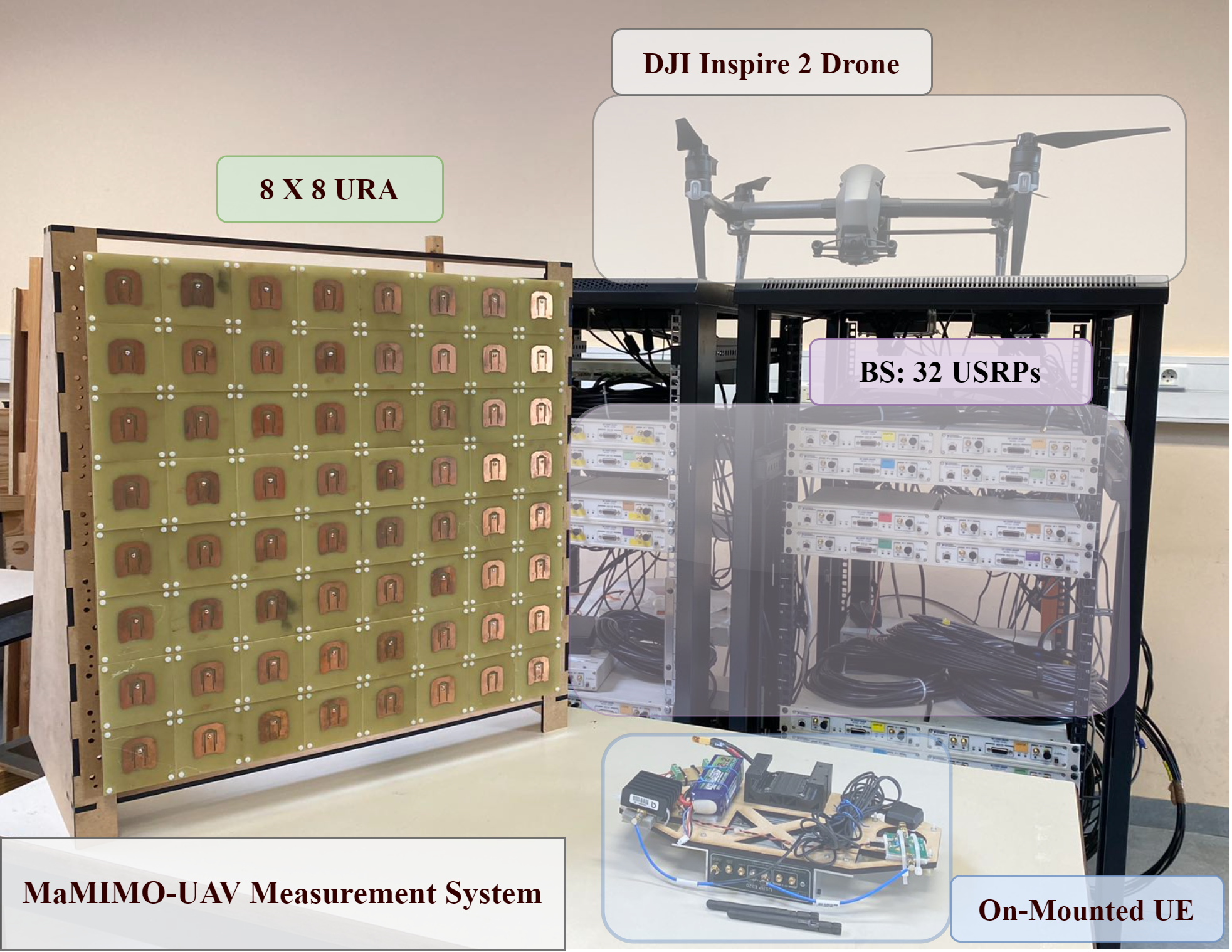}};
            \draw[black,ultra thick] (0.7,1.1) -- (0.51,5.05) node[midway, sloped, above, font=\small] {56~cm};
            \draw[black,ultra thick] (0.7,1.1) -- (4.05,1.9) node[midway, sloped, above, font=\small] {56~cm};
        \end{tikzpicture}
        \caption{MaMIMO UAV measurement system}
        \label{fig:hardware1}
    \end{subfigure}

    \begin{subfigure}{\linewidth}
        \centering
        \includegraphics[width=\textwidth]{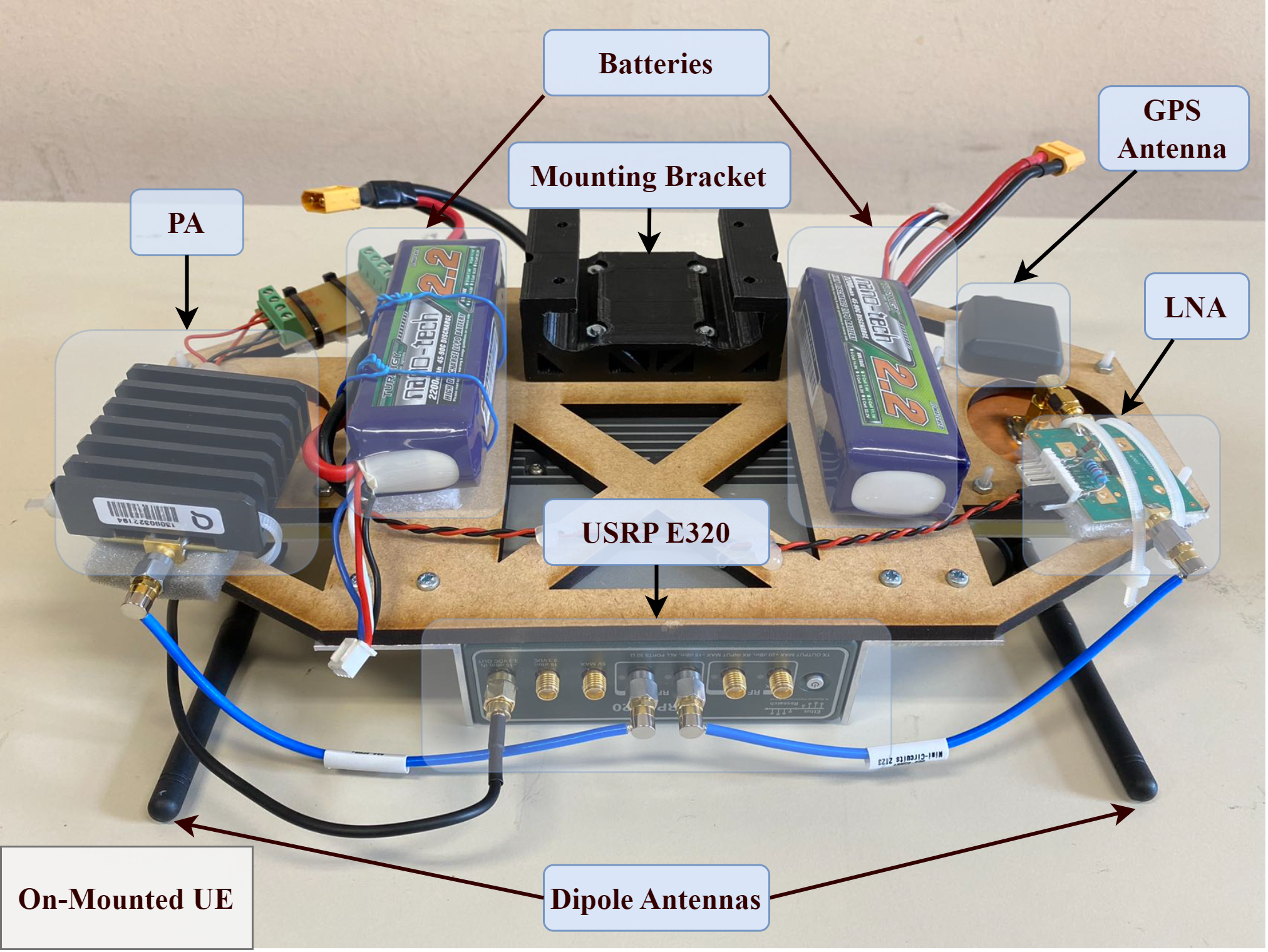}
        \caption{On mounted UE}
        \label{fig:hardware2}
    \end{subfigure}

	\caption{An overview of hardware in the A2G MaMIMO measurement system.}
	\label{fig:hardware}
\end{figure}

In this work we use a functional MaMIMO testbed to perform an OFDM-based A2G channel sounding using a fully connected MaMIMO antenna array.
The MaMIMO testbed consists of two main components, i.e., a multi-antenna BS and a single antenna user equipment (UE). 
%Thus, the Single-Input Multi-Output (SIMO) A2G channels can be captured.
The multi-antenna base station is made up of 32 USRPs, where each USRP has two transceiver channels. The MIMO BS connects to 64 patch antennas which can be composed into many different array configurations such as linear and planar antenna arrays. In this paper, we use a uniform rectangular array (URA) of $8\times8$ patch antennas as shown in Fig. \ref{fig:hardware1}. All USRPs use a hierarchy of octo-clock splitters which all use a single 10~MHz input reference clock provided by a single GPS Disciplined Oscillator (GPSDO). The hierarchy of octo-clock devices enables perfect synchronization between all 32 USRP's. The use of a GPSDO allows base station and user equipment to use the same reference input clock without the need to connect a coax cable between both sides.
In general, all USRPs handle all the baseband signal processing up to the radio frequency (RF) chain. The SIMO processing and some MAC layer processing of the BS are handled by FPGAs in the central system of the BS. The raw input and output bit streams are processed on a CPU and then send over a PCI interface to the FPGAs in the BS case or to the USRP directly in the UE case \cite{achiel23}.

The testbed operates with a TDD frame schedule where the BS provides synchronization signaling to the users. The BS performs channel estimation to conduct MIMO signal processing, such as precoding and combining. However, in the A2G measurement system, we focus on capturing the channel estimates, i.e., CSI, which allows for the characterization of the propagation channel.
The detailed method used by the BS to perform the channel estimation is as follows. First, the multi-antenna station broadcasts a synchronization signal to all the users. The user detects this synchronization and aligns its TDD frame to it. Next, the user transmits a user-specific uplink pilot sequence in the designated time slots. The BS will capture these uplink pilots and perform channel estimation based on the least-square (LS) algorithm for all users simultaneously. The system then uses the obtained CSI for precoding and combining. %\ac{refer algorithm?}\zc{check it is LS, MMSE, or something else?}, 
%where we received and stored the CSI. 

Modifications were made to the existing testbed to support A2G measurement campaigns: both on the software side and on the hardware side.

\subsubsection{Hardware}
To measure the A2G channel, the UE needs to be mounted on a drone. However, since the original UE is too bulky and heavy, a new UE was developed using a lightweight standalone USRP (E320). This device contains a Xilinx Zynq-7045 SoC together with an AD9361 RFIC. It is capable of implementing the necessary Over-the-Air (OTA) synchronization and the transmission of the uplink pilots. The casing is lightweight to be the payload of a UAV. The built-in ARM cores of the ZYNQ processor allow an embedded Linux to automatically configure and start the software and hardware on boot. The E320 features a built-in GPSDO allowing it to synchronize its internal clock oscillator to the same GPS clock as the BS.

To further improve the reliability of the OTA at the UE, a low noise amplifier (LNA), i.e., Qorvo QPA9807EVB-01, is connected to the receiving antenna port. A power amplifier (PA), i.e., Qorvo TQP9111-PCB2600, is connected to the transmission antenna port to increase the dynamic range of the measurement system. Both the USRP and PAs are powered by a LiPo battery, which allows powering the devices for a longer time than the total drone flight duration. Finally, the antennas used at the UE are dipole antennas. %The overview of the measurement system can be found in Table~I. %\zc{I suggest including a table to represent the measurement system, which can be together with measurement setup}.

The drone used in the measurements is a DJI~Inspire~2. A custom platform was designed to fit all the hardware. A custom-designed mounting bracket allowed the platform to be attached to the Inspire~2. An overview of the complete measurement system can be seen in Fig.~\ref{fig:hardware}. The total custom payload consists of a box containing the E320 USRP, a battery, a GPS antenna, a PA, an LNA, and 2 dipoles, as shown in Fig.~\ref{fig:hardware2}(b). The dipoles were pointed to the ground during the measurements.

\subsubsection{Software}
% The multi-antenna station FPGA code was modified such that the already existing channel estimations are extracted from the FPGA and are written back to a file. % Figure $ref figure$ shows the pipeline in the MIMO processing FPGA of the testbed where the channel estimation is performed. 
% The multi-antenna station performs a channel estimation every 0.5~ms, the channel estimates are temporarily stored in local DRAM because of the high data rate. When the required number of CSI samples is captured or when the DRAM is full, the measurement is stopped and all the estimates are written to a binary file on the hard drive. The capturing duration is thus limited by the amount of DRAM available, however, one can extend the capture period by sub-sampling the channel estimates and for example, saving channel estimate every 2~ms effectively quadrupling the available capture time.

%%% Second option

The testbed runs the massive MIMO framework in Labview NX 4.0 provided by NI. It is a fully customizable framework that allows modification of both the FPGA code and host side code. It implements an LTE-based Massive MIMO system that supports up to 12 users and 64 antennas. 
The system works in a TDD manner with predefined frame schedules.
The main frame symbols that are of interest are the PSS and the Uplink Pilots. The PSS is used for OTA synchronization in the downlink and the uplink pilots are used for uplink channel estimation. The LTE implementation takes care of the phase synchronization between user and base station.

To support channel measurements with a mobile user two main functions need to be implemented, first, the mobile user needs to synchronize its frame schedule to the BS and needs to send uplink pilots in the appropriate time slots, and second, the channel estimations in the downlink on the BS side need to be captured and written to a file.

\begin{figure}[t]
	\centering
    \includegraphics[scale=0.6]{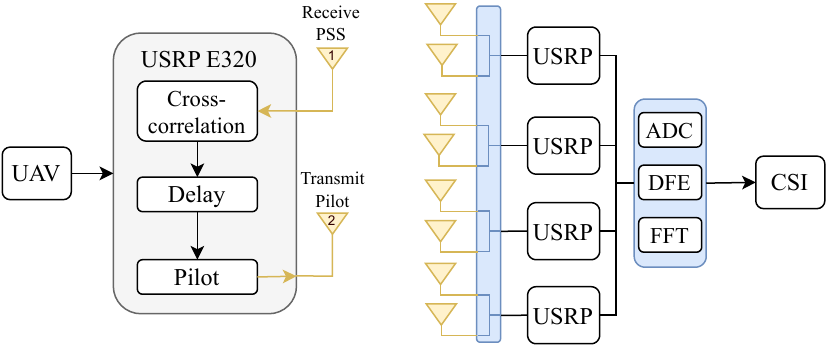}
	\caption{A block diagram of measured CSI collection process.}
	\label{fig:system_architecture}
\end{figure}

\textit{OTA Synchronization:}
The OTA synchronization is based on the Primary Synchronisation Symbol (PSS) in the downlink. The BS broadcasts the PSS at the start of each frame. The users aiming to synchronize are listening for the PSS. This feature was implemented in FPGA on the E320 and the full pipeline is shown in Fig.~\ref{fig:system_architecture}. To detect the PSS a complex cross-correlation with the original PSS signal is performed. Following the cross-correlation a peak detector is implemented to detect the start of each frame. Once the start of a frame is known, the UE waits the appropriate delay time to send a repetition of pilot sequences in the correct frame schedule slots.

\textit{Channel Capture:}
An extension of the LabVIEW MaMIMO framework has been implemented that allows the real-time capturing of the channel state estimations calculated at BS side. Fig.~\ref{fig:system_architecture} shows the first stages of the pipeline in the MIMO processing FPGA of the testbed up to the channel estimation. The testbed performs channel estimation resulting in a $\textbf{H}(t,\Delta f)$ channel frequency response vector as defined in Section \ref{sec:channel} of dimensions $64\times1$ for 100 frequency bins for each user. Every element of the vector is a 16-bit fixed point complex number and since the channel is estimated every 0.5~ms this generates a bit rate of approximately 410~Mbps. The channel estimations are thus saved to DRAM during the measurement and afterward written to a binary file. The main limitation to the maximum measurement duration is thus the size of DRAM storage, however, the measurement duration can be extended by subsampling channel estimations written to DRAM, for example by saving a channel estimate every 1~ms the maximum measurement duration is effectively doubled.

\subsection{Measurement Scenario}
\begin{figure}
	\centering
 
    % \begin{scaletikzpicturetowidth}{\linewidth}
    \input{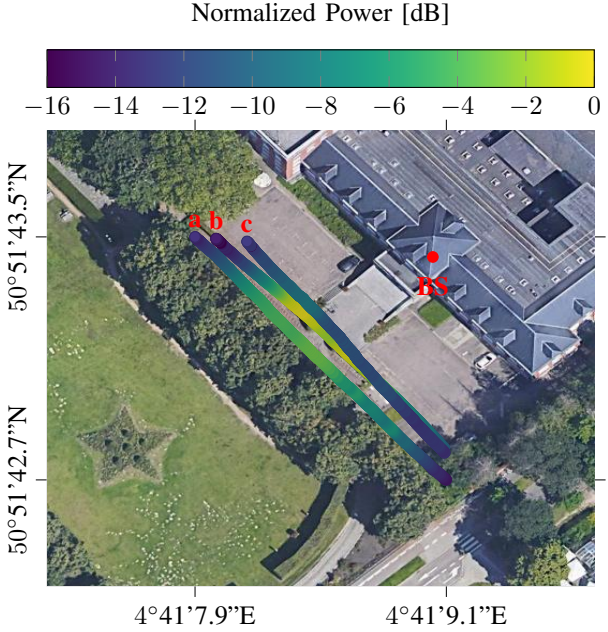}
    % \end{scaletikzpicturetowidth}
	\caption{Normalized received power results for $a$, $b$, and $c$ trajectories corresponding to 8, 11, and 24~m, respectively. The red dot represents the location of the square antenna array at an altitude of 11~m, image source@Google Map.}%\ac{check whether we are allowed to use google satellite for IEEE}}
	\label{fig:prx_vs_dist}
\end{figure}

To showcase the complete measurement system and the potential for A2G channel modeling, a suburban campus scenario is selected and measured, where the main scatterers include buildings, trees, and the ground plane. In section \ref{sec:results} we discuss the results of the measurement in this scenario, where the picture of the scenario is seen in Fig.~\ref{fig:prx_vs_dist}. 
The target of the measurement scenario is to emulate a drone flying at different altitudes above ground level in a straight line while being connected to a BS on the ground. To emulate this, the antenna array of the base~station is put in a window facing outwards at an altitude of 11~m. The UAV flies following a trajectory parallel to the BS antenna array. The same trajectory is repeated at three different altitudes: 8~m, 11~m, and 24~m. The height of 11~m was chosen as this is the same as the altitude of the base station antenna array. A lower height of 8~m was chosen as this was the lowest practically possible altitude taking into account necessary safety measures. Finally, the height of 24~m was chosen as this the same as the rooftop height of the nearby building. All trajectories can be seen in Fig.~\ref{fig:prx_vs_dist}. During the measurements the center frequency was 2.61~GHz, the bandwidth was 18~MHz, transmit power was 30~dBm, and the receive gain at the BS was fixed to 15~dB. The UAV transmits using an interval transmission a standard LTE compliant sounding signal every 1~ms. It is an OFDM signal consisting out of a Zadoff-Chu sequence. During each measurement, the GPS location of the UAV was logged every 10~ms. The drone flew at a constant speed of 1.5~m/s in a straight line. All of the sampling locations were in the line of sight (LOS) of the antenna array of the BS. Both the BS and the user used a GPS-disciplined oscillator as their input reference clock to minimize frequency offset. All measurement parameters can be found in Table~\ref{tab:params}.

\begin{table}[]
\centering
\caption{Parameters used during the measurement} \label{tab:params}
\begin{tabular}{ll}
\textbf{Parameter}   & \textbf{Value} \\ \hline
UAV heights              & 8~m, 11~m, 24~m   \\ \hline
BS antenna array height & 11~m   \\ \hline
Center frequency     & 2.61~GHz        \\ \hline
Bandwidth            & 18~MHz          \\ \hline
Speed                & 1.5~m/s         \\ \hline
CSI logging interval & 1~ms            \\ \hline
GPS logging interval & 10~ms           \\ \hline
\end{tabular}
\end{table}

\section{Results and Evaluations}
\label{sec:results}
In this section, we present the %numerous \zh{or numerical?} 
results obtained from the measurement data. We first provide some general results including PDPs and received power. Then, 3D channel correlation and corresponding stationarities are evaluated. 

\begin{figure}
	\centering
    \begin{subfigure}{\linewidth}
        \centering
        \includegraphics[width=0.8\textwidth]{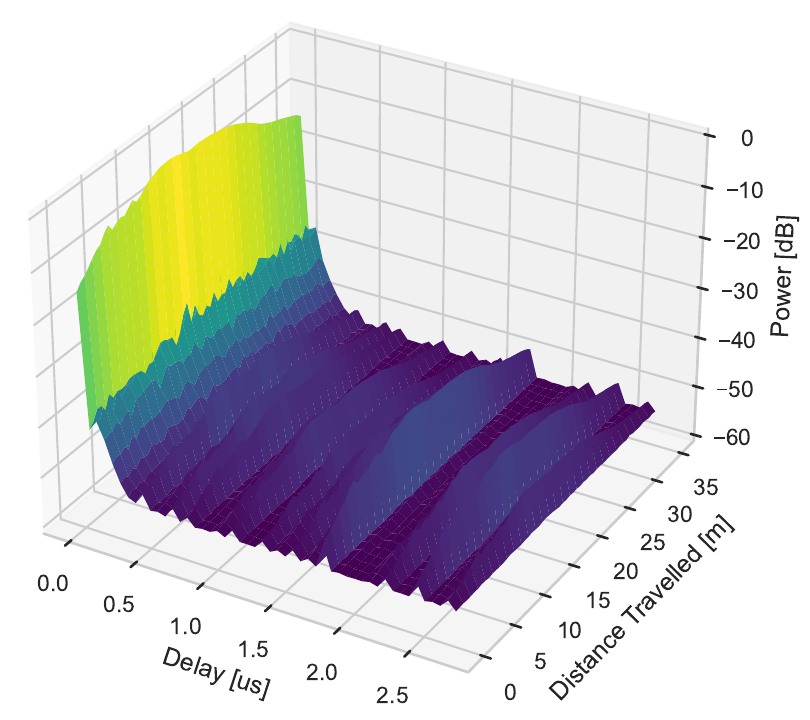}        
        \caption{Height: 8~m}
    \end{subfigure}
    \medskip
    \begin{subfigure}{\linewidth}
        \centering
        \includegraphics[width=0.8\textwidth]{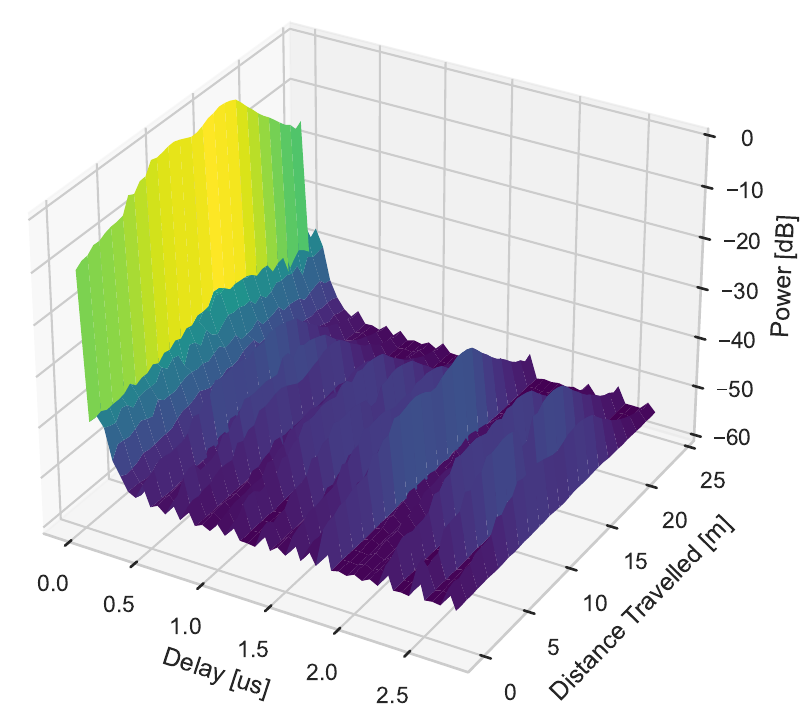}
        \caption{Height: 11~m}
    \end{subfigure}
    \medskip 
    \begin{subfigure}{\linewidth}
        \centering
        \includegraphics[width=0.8\textwidth]{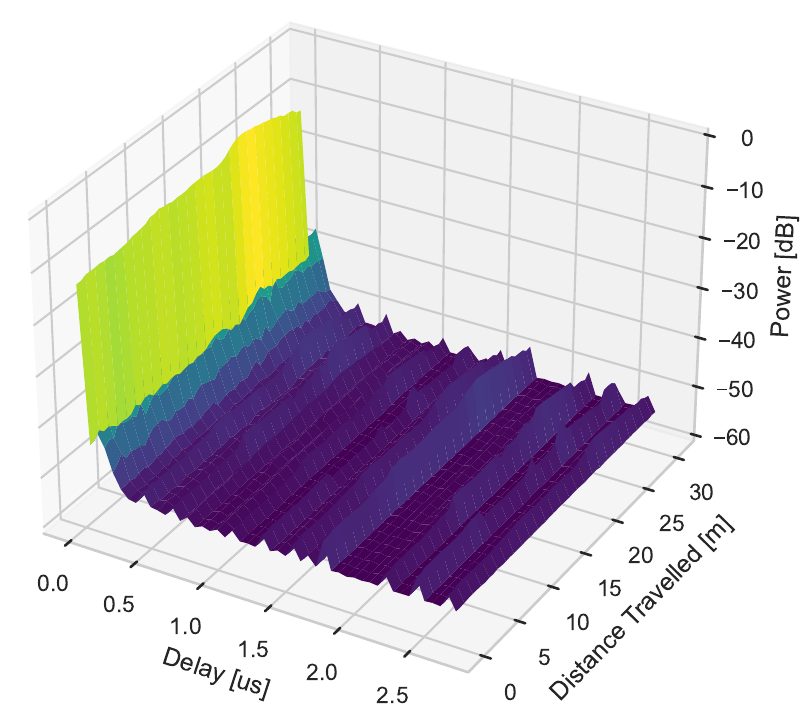}
        \caption{Height: 24~m}
    \end{subfigure}
	\caption[Normalized averaged PDPs for different heights and traveling distances.]{Normalized averaged Power Delay Profiles for different heights and traveling distances.}
    \label{fig:avg_pdp}
\end{figure}
\subsection{Received Power and PDP}
The normalized received power results are shown in Fig.~\ref{fig:prx_vs_dist}. The three trajectories at different altitudes are shown as $a$, $b$, and $c$ representing altitudes 8~m, 11~m, and 24~m respectively. The location of the rectangular antenna array of the BS at an altitude of 11~m is indicated as a red dot. Fig.~\ref{fig:prx_vs_dist} also shows the GPS coordinates within which the measurement has taken place.
As expected the received power at the locations close to the BS is higher than at the edges of the trajectories. Similarly, the received power at 24~m is lower than at 11~m due to the increased distance.

The resulting normalized averaged PDPs for the three altitudes can be seen in Fig.~\ref{fig:avg_pdp}. The PDP $\textbf{P} (d,\tau)$ is obtained from Eq.~(\ref{eq:pdp}) with the traveling distance of $d=vt$ where the speed of UAV $v$ is nearly constant (approximately $1.5~m/s$). We can observe that the PDP at 11~m has larger peaks at different delay times than the other two PDPs, which indicates more obvious multipath effects are present at this altitude. 
In addition, the LOS power for the relative delay at 0 is the received power in Fig.~\ref{fig:prx_vs_dist}, which shows an increasing difference in the order of 24~m, 8~m, and 11~m, with the maximum power of -7.5~dB, -2.9~dB, and 0~dB, respectively.%, which suggests the UAV flying at the same height level as the BS can gain the higher power. %\ac{isn't this a bit redundant?} \zc{it is ok, otherwise it is too short description} 

% \begin{figure}
% 	\centering
%     \includegraphics[scale=0.5]{img/comb_cmd.png}
% 	\caption{CMDs based on angle to the BS and traveling distance for three different altitudes.}
% 	\label{fig:cmd}
% \end{figure}

\begin{figure*}
	\centering
    \begin{scaletikzpicturetowidth}{0.85\linewidth}
        % This file was created with tikzplotlib v0.10.1.
\begin{tikzpicture}

\definecolor{darkslategray38}{RGB}{38,38,38}
%\definecolor{white!15!black}{RGB}{204,204,204}

\begin{groupplot}[group style={group size=3 by 2, vertical sep=2cm}]
\nextgroupplot[
scale=\tikzscale,
axis equal image,
axis line style={white!80!black},
tick align=outside,
title={Height 8 m},
x grid style={white!80!black},
xtick pos=left,
ytick pos=left,
xmajorgrids,
xmajorticks=true,
xmin=40, xmax=140,
xtick style={color=black},
y grid style={white!80!black},
ylabel=\textcolor{darkslategray38}{Angle to basestation [deg]},
ymajorgrids,
ymajorticks=true,
ymin=40, ymax=140,
ytick style={color=black}
]
\addplot graphics [includegraphics cmd=\pgfimage,xmin=41.4960615123498, xmax=139.005007232007, ymin=41.4960615123498, ymax=139.005007232007] {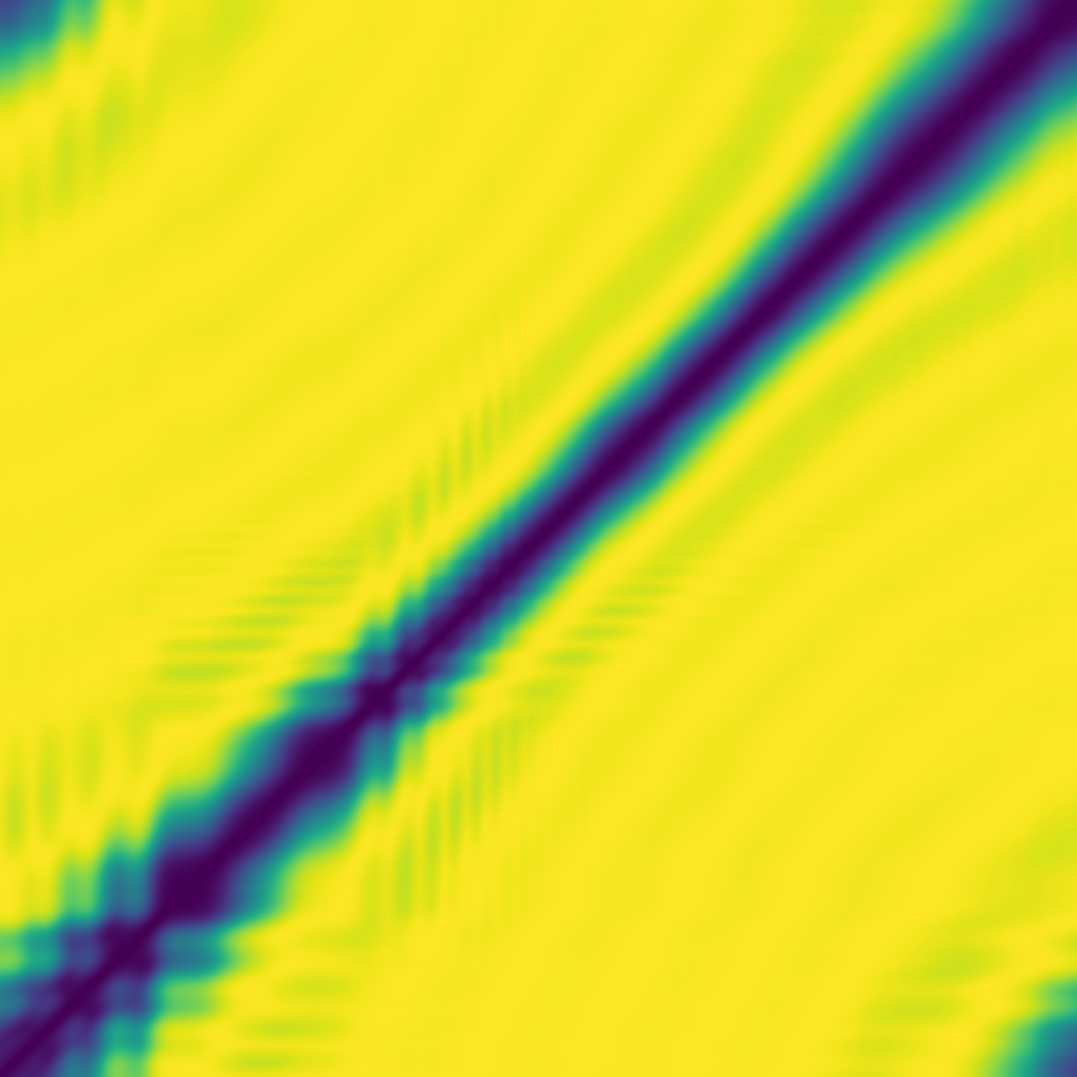};

\nextgroupplot[
    scale=\tikzscale,
    axis equal image,
axis line style={white!80!black},
tick align=outside,
title={Height 11 m},
x grid style={white!80!black},
xlabel=\textcolor{darkslategray38}{Angle to basestation [deg]},
xtick pos=left,
ytick pos=left,
xmajorgrids,
xmajorticks=true,
xmin=40, xmax=140,
xtick style={color=black},
y grid style={white!80!black},
ymajorgrids,
ymajorticks=false,
ymin=40, ymax=140,
ytick style={color=black}
]
\addplot graphics [includegraphics cmd=\pgfimage,xmin=41.4986051494578, xmax=139.00738896783, ymin=41.4986051494578, ymax=139.00738896783] {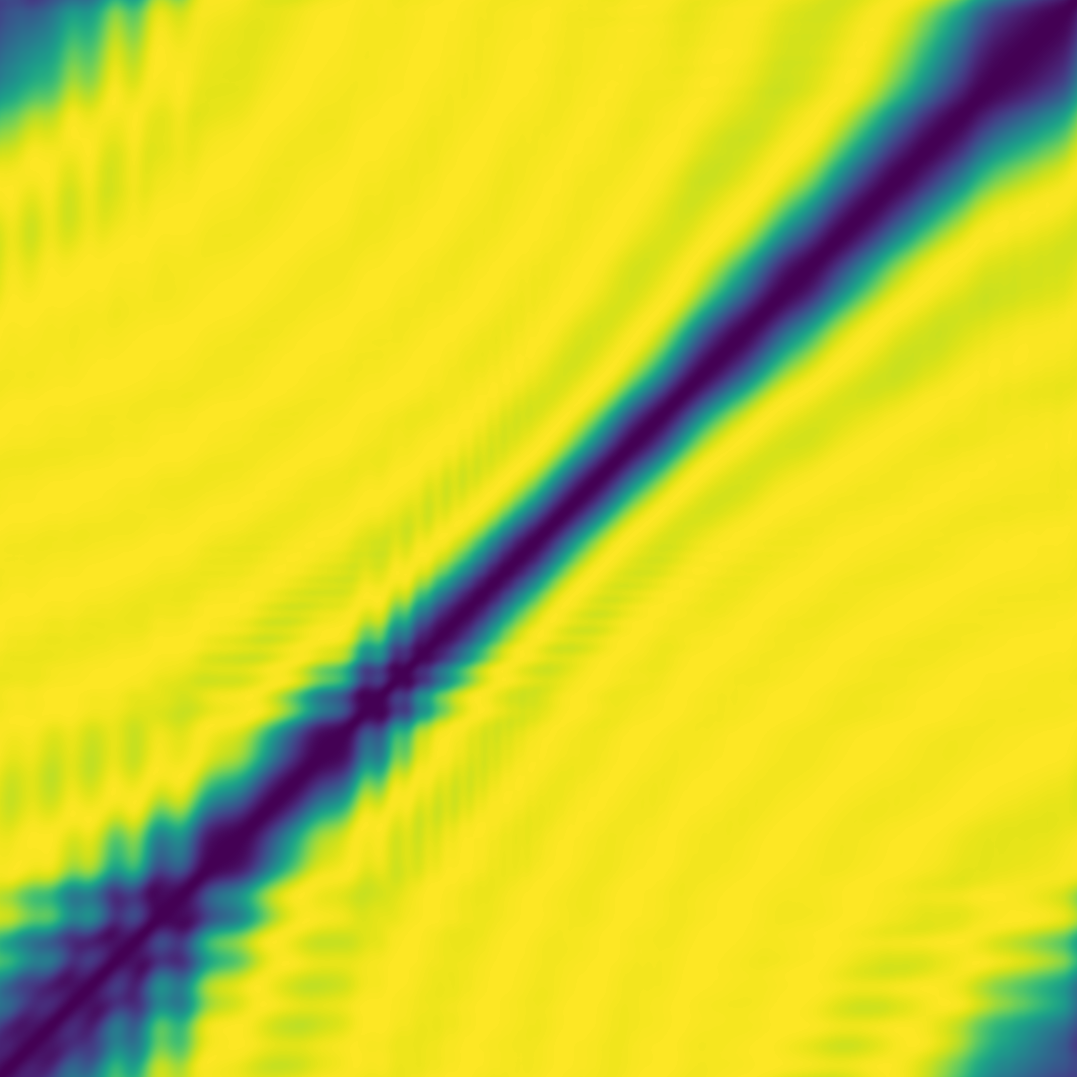};

\nextgroupplot[
    scale=\tikzscale,
    axis equal image,
axis line style={white!80!black},
tick align=outside,
title={Height 24 m},
x grid style={white!80!black},
xmajorgrids,
xmajorticks=true,
xmin=40, xmax=140,
xtick pos=left,
ytick pos=left,
xtick style={color=black},
y grid style={white!80!black},
ymajorgrids,
ymajorticks=false,
ymin=40, ymax=140,
ytick style={color=black},
colorbar right,
    every colorbar/.append style={height=
        2*\pgfkeysvalueof{/pgfplots/parent axis height}+
        \pgfkeysvalueof{/pgfplots/group/vertical sep}},
colorbar style={ylabel={CMD, $d_{corr}$}},
colormap/viridis,
point meta max=1,
point meta min=0
]
\addplot graphics [includegraphics cmd=\pgfimage,xmin=41.4967063867429, xmax=139.003325871051, ymin=41.4967063867429, ymax=139.003325871051] {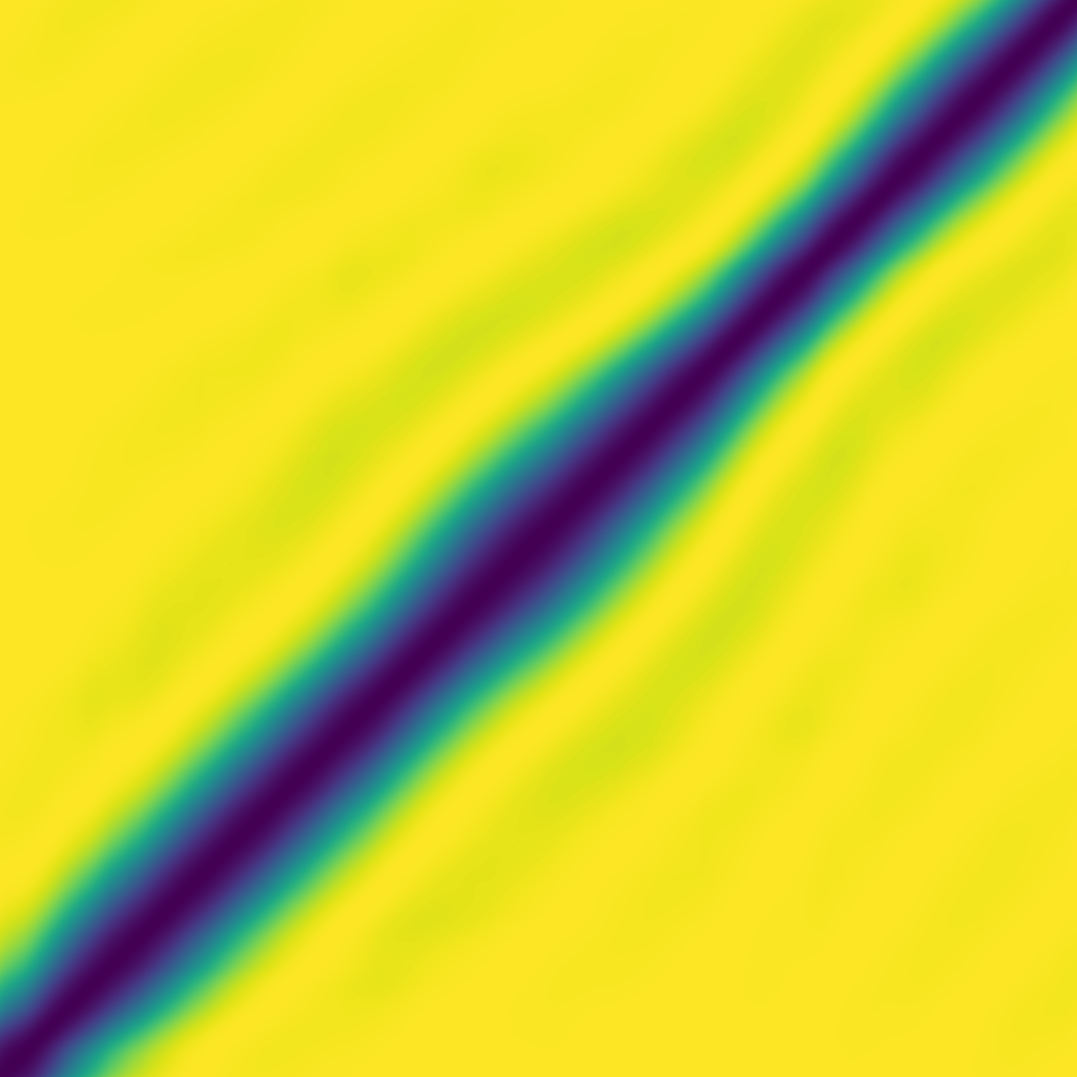};

\nextgroupplot[
    scale=\tikzscale,
    axis equal image,
axis line style={white!80!black},
tick align=outside,
x grid style={white!80!black},
xmajorgrids,
xmajorticks=true,
xmin=0, xmax=30.2476546901156,
xtick style={color=black},
xtick pos=left,
ytick pos=left,
y grid style={white!80!black},
ylabel=\textcolor{darkslategray38}{Distance travelled [m]},
ymajorgrids,
ymajorticks=true,
ymin=0, ymax=30.2476546901156,
ytick style={color=black}
]
\addplot graphics [includegraphics cmd=\pgfimage,xmin=0, xmax=33.6868035406563, ymin=0, ymax=33.6868035406563] {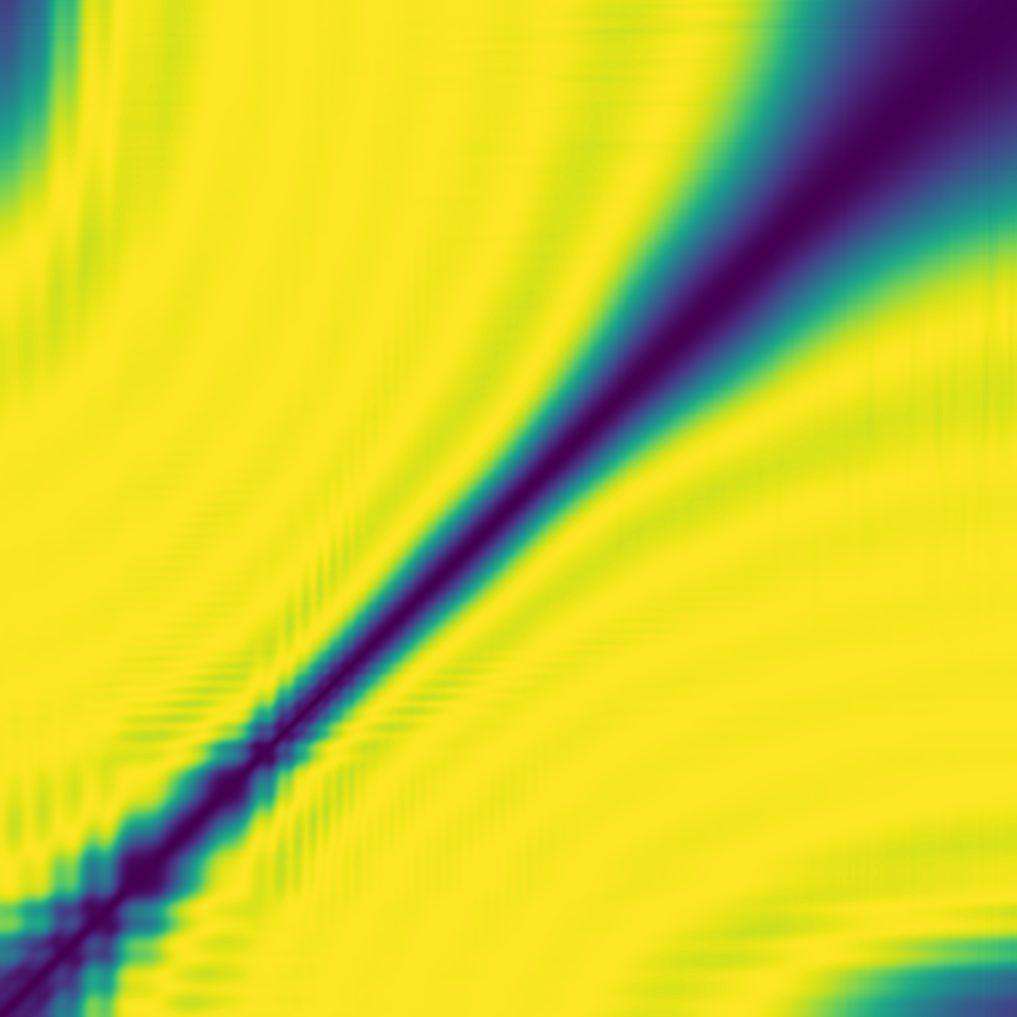};

\nextgroupplot[
    scale=\tikzscale,
    axis equal image,
axis line style={white!80!black},
tick align=outside,
x grid style={white!80!black},
xlabel=\textcolor{darkslategray38}{Distance travelled [m]},
xtick pos=left,
ytick pos=left,
xmajorgrids,
xmajorticks=true,
xmin=0, xmax=30.2476546901156,
xtick style={color=black},
y grid style={white!80!black},
ymajorgrids,
ymajorticks=false,
ymin=0, ymax=30.2476546901156,
ytick style={color=black}
]
\addplot graphics [includegraphics cmd=\pgfimage,xmin=0, xmax=30.2476546901156, ymin=0, ymax=30.2476546901156] {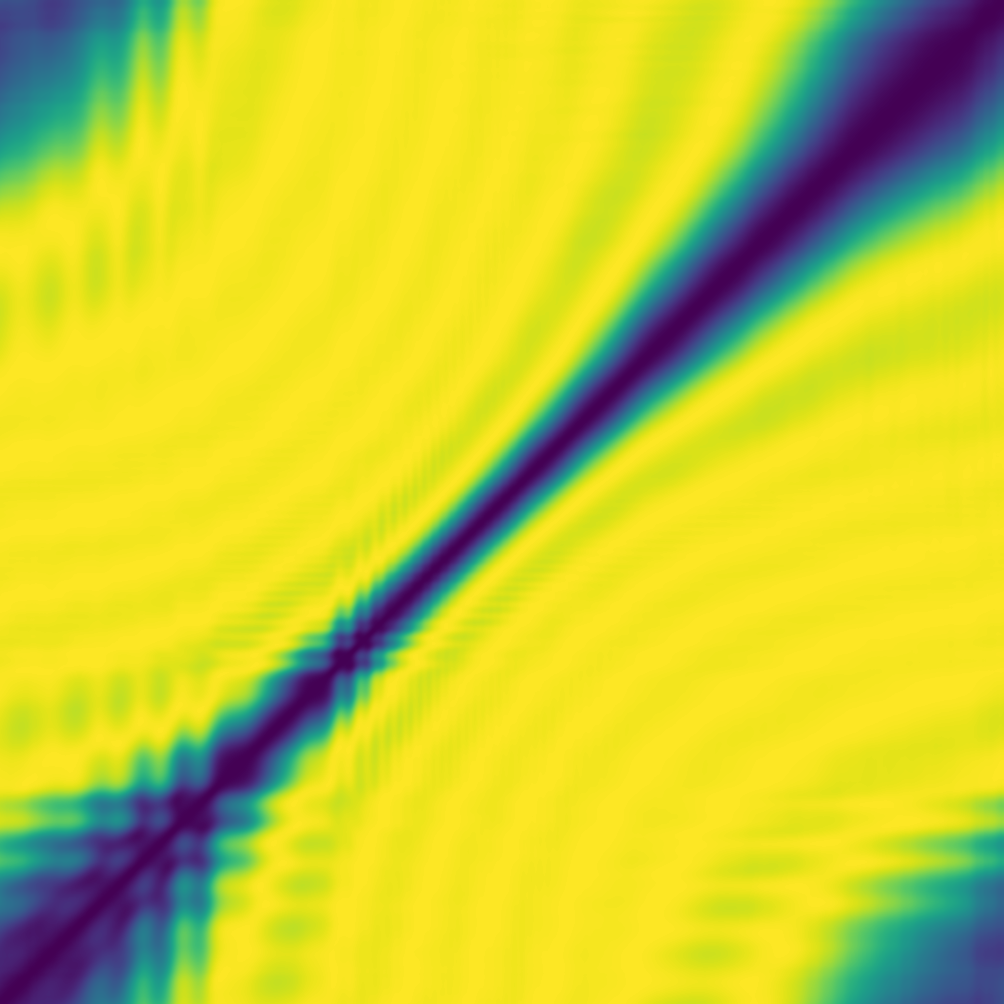};

\nextgroupplot[
    scale=\tikzscale,
    axis equal image,
axis line style={white!80!black},
tick align=outside,
x grid style={white!80!black},
xmajorgrids,
xmajorticks=true,
xmin=0, xmax=30.2476546901156,
xtick style={color=black},
xtick pos=left,
ytick pos=left,
y grid style={white!80!black},
ymajorgrids,
ymajorticks=false,
ymin=0, ymax=30.2476546901156,
ytick style={color=black}
]
\addplot graphics [includegraphics cmd=\pgfimage,xmin=0, xmax=30.6237070977718, ymin=0, ymax=30.6237070977718] {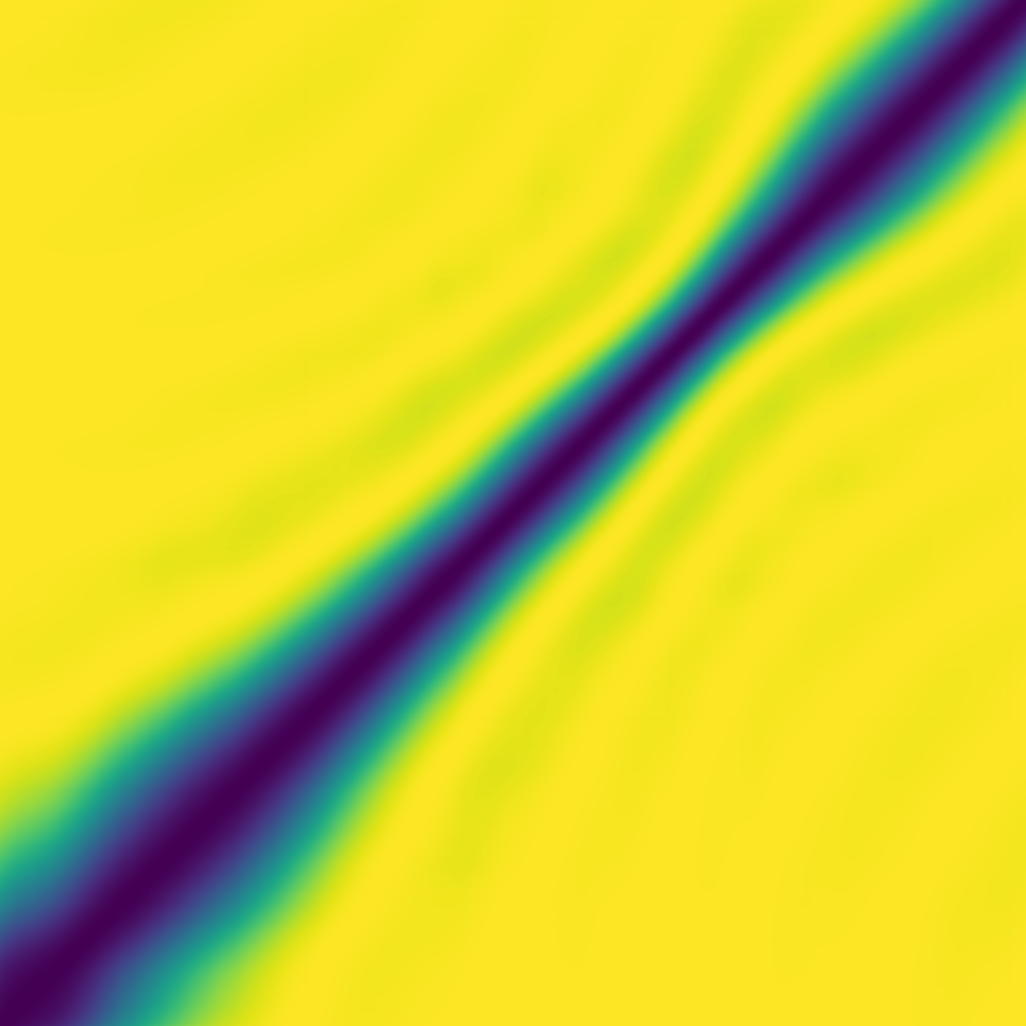};
\end{groupplot}

\end{tikzpicture}
    \end{scaletikzpicturetowidth}
    \centering
	\caption{CMDs based on angle to the base station and traveling distance for three different altitudes.}
	\label{fig:cmd}
\end{figure*}

\subsection{Temporal Stationarity}
\label{sec:sa}
Same with the PDP processing, considering the constant velocity of the drone, the time stationarity can be characterized by the distance, where the quasi-stationary distance (QSD) is a popular metric in existing works \cite{he2015characterization,cui2018acceleration}. As an indispensable step to obtain QSD, we first calculate the CMD based on Eq.~(\ref{eq:cmd}), as shown in the second row of Fig.~\ref{fig:cmd}. 

\begin{figure}
    \centering 
    \includegraphics[width=0.9\linewidth] {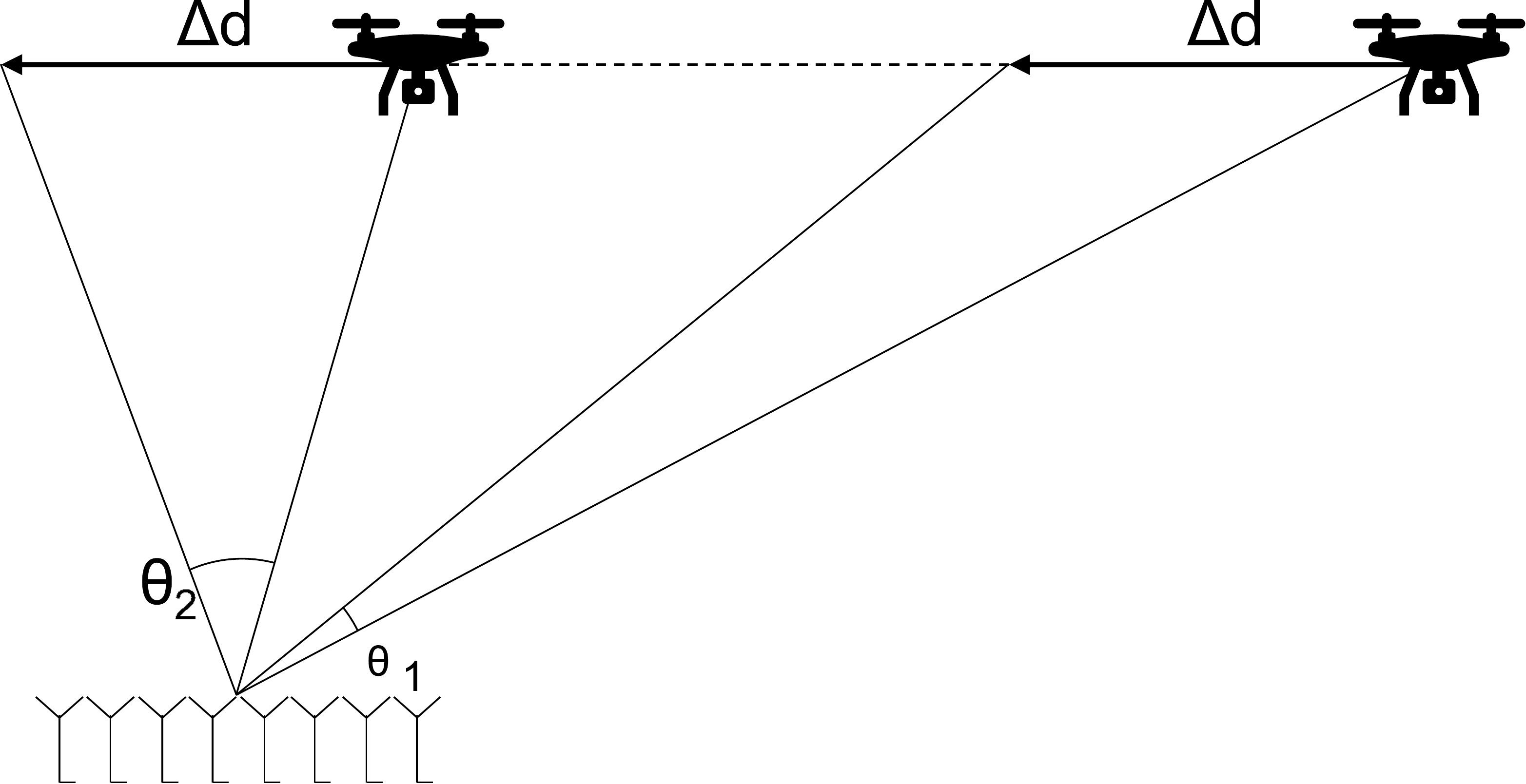}
    \caption{A schematic representation of a user travelling %a fixed distance $\Delta d$ 
    along a straight trajectory parallel to BS antenna array, where the drone flying the same $\Delta d$ when close to and far away the antenna array will result in a larger $\theta_2$ and a smaller $\theta_1$, respectively.}
    \label{fig:angles}
\end{figure}

A narrow region of low CMD values along the diagonal indicates that the correlation changes rapidly along the traveling distance resulting in a short stationary distance. However, at the start and the end of the trajectory, the difference in correlation becomes smaller leading to a more wide region of low CMD values. This wider region would imply a larger stationary distance at the start and end of the trajectory although the environment is similar to the center of the trajectory and the speed of the UAV constant. This can be explained by the fact that when the UAV travels a set distance $\Delta d$ in the center of the trajectory the angle between the UAV and all antenna elements of the URA changes more rapidly than when the UAV travels the same set distance $\Delta d$ at the start or end of the trajectory as shown in Fig. \ref{fig:angles}.

\textit{Angular Stationarity:}
Since the location of the base station and the user is always known, we can convert traveling distance to traveling angle using simple geometry. Next, we consider the angle to BS as the reference to quantify time stationarity instead of the traveled distance. In reality, the angles of departure or arrival are important metrics to evaluate MIMO systems, moreover, channel models based on angle estimation have been proposed in \cite{jiangaoa}. Here, We define the stationary angle as the spatial angle over which the stationarity holds. Angular stationarity is especially relevant in MaMIMO scenarios where a large number of antennas is present on one side since it is the variation of the angle to the antenna array that will have a large impact on the channel to each antenna element separately. Angular stationarity is an important metric in MaMIMO system design for optimizing beam tracking algorithms as it gives a guideline on how frequently a new beam should be assigned to a moving user based on angular estimation.
In this work, the location at the boresight of the antenna array is considered at an azimuth angle of $90^{\circ}$. The data of the trajectories is limited to the interval of $40-140^{\circ}$ to be able to compare the different altitudes in a fair manner. The step size of the angle is $0.01^{\circ}$.
Note that, because of the geometric relationship, for an equal angle, the traveling distance becomes different.

As a result, we can see that the CMD becomes more stable over the whole trajectory, as shown in the three sub-figures on the top of Fig.~\ref{fig:cmd}. Moreover, for different heights of the UAV, it shows the CMD at 11~m presents more obvious changes than that at the other two heights. With the obtained CMD, we can calculate the stationary distance and angle using Eq.~(\ref{eq:sd}).  

\begin{figure}
	\centering
        \input{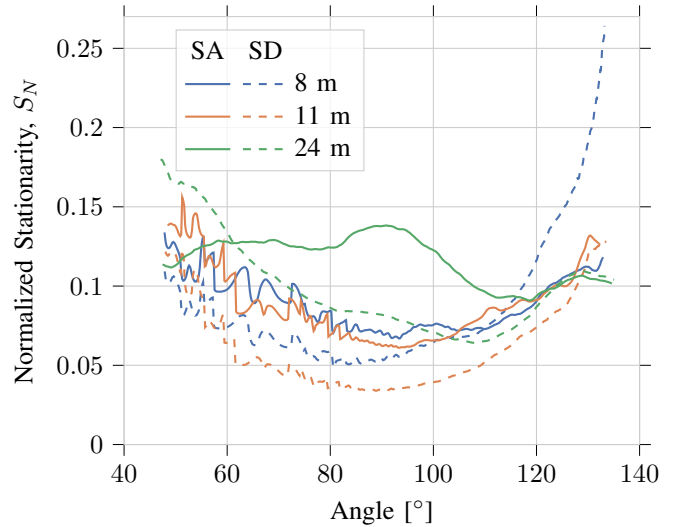}
    \caption{Normalized time stationarity for both distance based and angle based stationarity with respect to angle to the BS for different altitudes.}
    \label{fig:stationarity_time}
\end{figure}
Firstly, the absolute values of the obtained stationary distance/angle make it difficult to compare the results of both metrics. Thus, we introduce a new normalized stationarity metric, equal to the ratio between the obtained stationary distance/angle and the total traveling distance/angle. For the stationary distance it is calculated as follows:
\begin{equation}
    S_N =\frac{d_{SD}}{d_{total}}
\end{equation}
where $d_{SD}$ is the stationary distance in meters and $d_{total}$ is the total distance traveled in meters. For the stationary angle it is calculated as follows:
\begin{equation}
    S_N=\frac{d_{SA}}{a_{total}}
\end{equation}
where $d_{SA}$ is the stationary angle in degrees and $a_{total}$ is the total angle in degrees covered from the start to the end of the trajectory. The normalized stationarity (in percentage) allows us to compare the stationary distance and stationary angle values since the total trajectory is the same for both. It represents the fraction of the total trajectory over which the channel can be considered stationary.

%provides a more reasonable reference to UAVs with different trajectories. 
The results are shown in Fig.~\ref{fig:stationarity_time}. It can be found that the normalized stationary angle and distance at the height of 11~m are generally lower than the other two heights, which can be explained by the largest power variation at this height, as shown in Fig.~\ref{fig:avg_pdp}(b). Fig.~\ref{fig:stationarity_time} shows that for example at a height of 11~m the normalized stationary distance clearly increases from the center of the trajectory  with a value of $0.04$ to a value of $0.12$ at the edge of the trajectory. This phenomenon can be observed at all trajectory altitudes. In comparison, the normalized stationary angle stays more constant over the whole trajectory. Overall, the normalized stationarity is below 15\%, where the low values suggest the high dynamics of A2G channels.
%Besides, the normalized stationary distance at 8~m dramatically increases when the UAV flew far away from the BS ($\phi>110^{\circ}$) because ??. 
%\zc{Overall, the normalized stationarity is below 15\%, which means that the A2G channel becomes outdated...\zc{more description.}}

\begin{figure}[t]
	\centering
    \input{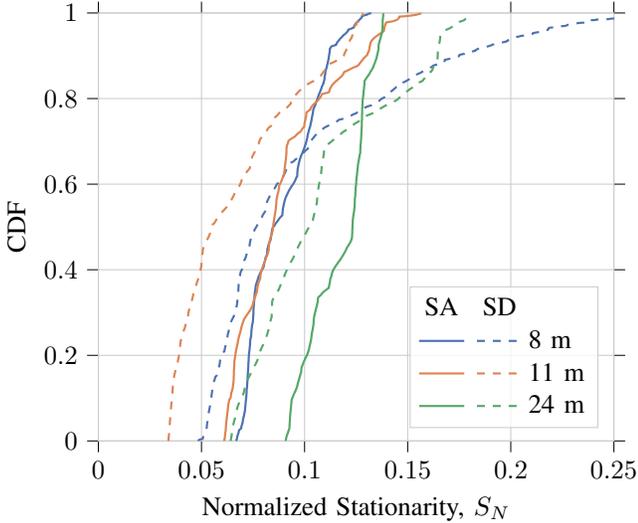}
	\caption{CDFs of normalized time stationarity for both distance based and angle based stationarity at different altitudes.}
	\label{fig:CDF_st}
\end{figure}

To compare the difference among heights and metrics (angle and distance), a more intuitive way is to compare the statistical CDF. As shown in Fig.~\ref{fig:CDF_st}, it can be seen that the mean of the stationary distance is lower than the mean of the stationary angle for the same altitude trajectory. Also, the steepness of the CDF gives a clear indication that the values of the stationary angle are more stable than the stationary distance along the same trajectory. The absolute values of the mean ($\mu$) and standard deviation ($\sigma$) for the stationary angle at 8~m, 11~m, and 24~m are $8.8^{\circ}$/$1.6^{\circ}$($\mu/\sigma$), $8.7^{\circ}$/$2.3^{\circ}$, and $11.5^{\circ}$/$1.4^{\circ}$ respectively and can be found in Table~\ref{tab:stat}, which shows that for the higher-altitude drone's trajectory, the channel stationarity is more stable and that the stationary angle is a better metric since we would expect stable time stationarity when the propagation environment does not change drastically along the trajectory. More statistical results can be found in Table~\ref{tab:stat}.

%\zc{table needed}.
\setlength\tabcolsep{5pt}
\begin{table}[tbp]
\centering
  \caption{Statistical Results of Key Parameters } \label{tab:stat}
  \begin{tabular}{c|c|c|c|c|c|c|c|c}
\hline
  \textbf{Trajectory} & \multicolumn{2}{|c|}{\textbf{SD} [m]} & \multicolumn{2}{|c|}{\textbf{SA} [deg]}  & \multicolumn{2}{|c|}{\textbf{$S_{\tau}$ [ns]}} & \multicolumn{2}{|c}{\textbf{$B_{coh}$ [MHz]}} \\
    \hline
  Height [m] & $\mu$ & $\sigma$ & $\mu$  & $\sigma$ & $\mu$  & $\sigma$ & $\mu$  & $\sigma$ \\
  \hline
8 & $3.3$ & $1.7$ & $8.8$ & $1.6$ & $405.4$ & $113.2$ & $10.6$ & $1.2$ \\
  \hline
11 & $2.2$ & $1.0$ & $8.7$ & $2.3$ & $341.6$ & $138.9$ & $10.9$ & $1.0$ \\
  \hline
24 & $3.6$ & $1.1$ & $11.5$ & $1.4$ & $454.9$ & $43.6$ & $8.7$ & $0.1$ \\
  \hline
\end{tabular}
\end{table}

\subsection{Frequency Stationarity}
As mentioned in Section II, the frequency stationarity can be observed in two ways: one way is by looking at the RMS delay spread calculated by PDP, and the other is by directly calculating the correlation of the channel frequency response and then obtaining the coherent bandwidth using a threshold of $1/e$.
The coherence bandwidth $B_{coh}$ and root mean square delay spread $S_{\tau}$ are interconnected, as $S_{\tau}$ is derived from the PDP. In contrast, $B_{coh}$  is obtained from the frequency correlation function, which is the Fourier transform of the PDP. Nonetheless, this relationship is typically ambiguous and contingent upon specific system configurations and environmental factors \cite{molish2011wireless}. To signify this relationship, we introduce the parameter $\alpha$. In the context of the corresponding UAV system, the determined values of $\alpha$ can prove valuable for various system parameters of interest, such as frequency-division or time-division systems, as it converts time-frequency relationships.
The relationship between them can be empirically expressed as $B_{coh}=\frac{1}{\alpha S_\tau}$, where $\alpha$ is the scaling parameter. From the mean values in Table~I, we can find $\alpha=$ 0.23, 0.27, and 0.25 for trajectories at 8, 11, and 24~m, respectively. The similar values on the one hand indicate the same propagation environment, on the other hand, verify the correctness of channel analysis.

\begin{figure}[t]
	\centering
    \input{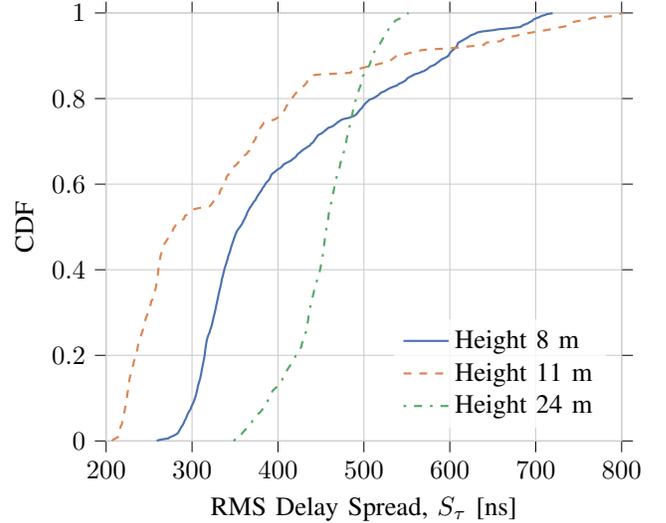}
	\caption{CDFs of RMS delay spread for different altitudes.}
	\label{fig:rms_delay}
\end{figure}

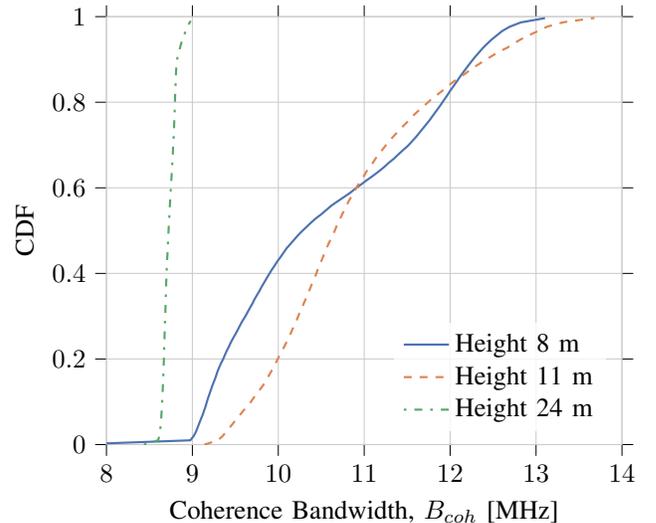
\begin{figure}[t]
	\centering
    % This file was created with tikzplotlib v0.10.1.
\begin{tikzpicture}

%\definecolor{color2}{RGB}{85,168,104}
%\definecolor{color1}{RGB}{221,132,82}
%\definecolor{color0}{RGB}{76,114,176}

\begin{axis}[
axis line style={white!80!black},
tick align=outside,
tick pos=both,
x grid style={white!80!black},
xmajorgrids,
xtick style={color=black},
y grid style={white!80!black},
ymajorgrids,
ytick style={color=black},
legend cell align={left},
legend pos=south east,
legend style={fill opacity=0.8, draw opacity=1, text opacity=1, draw=none},
ymin=0, ymax=1,
ylabel={CDF},
xmin=8, xmax=14,
xlabel={Coherence Bandwidth, $B_{coh}$ [MHz]},
]
\addplot [thick, color0]
table {%
7.61714267730713 0
8.97571468353271 0.0099647045135498
9.01571464538574 0.0199292898178101
9.041428565979 0.0298939943313599
9.06142902374268 0.0398585796356201
9.07999992370605 0.0498232841491699
9.11999988555908 0.06975257396698
9.1542854309082 0.08968186378479
9.16857147216797 0.0996465682983398
9.18428611755371 0.1096111536026
9.23571395874023 0.13950514793396
9.25571441650391 0.14946985244751
9.27428531646729 0.15943443775177
9.28999996185303 0.16939914226532
9.3100004196167 0.17936384677887
9.33142852783203 0.18932843208313
9.35714244842529 0.19929313659668
9.39999961853027 0.21922242641449
9.42571449279785 0.22918701171875
9.44857120513916 0.2391517162323
9.47571468353271 0.24911642074585
9.50428581237793 0.25908100605011
9.55571460723877 0.27901029586792
9.58285713195801 0.28897500038147
9.61428546905518 0.29893970489502
9.66857147216797 0.31886899471283
9.69714260101318 0.32883358001709
9.72142887115479 0.33879828453064
9.75 0.348762989044189
9.77999973297119 0.35872757434845
9.80571460723877 0.368692278862
9.83571434020996 0.37865686416626
9.86428546905518 0.38862156867981
9.92428588867188 0.40855085849762
9.958571434021 0.418515563011169
9.98999977111816 0.42848014831543
10.0271425247192 0.438444852828979
10.0957145690918 0.45837414264679
10.1828575134277 0.4783034324646
10.2257146835327 0.488268136978149
10.2742853164673 0.49823272228241
10.3785715103149 0.51816201210022
10.4342861175537 0.52812671661377
10.5 0.538091421127319
10.6157140731812 0.558020710945129
10.6871433258057 0.56798529624939
10.834285736084 0.587914705276489
10.8957147598267 0.59787929058075
10.9642858505249 0.607843995094299
11.0299997329712 0.61780858039856
11.1000003814697 0.627773284912109
11.165714263916 0.637737989425659
11.2271432876587 0.647702574729919
11.2814283370972 0.65766716003418
11.3428573608398 0.667631864547729
11.4485712051392 0.687561273574829
11.5057144165039 0.697525858879089
11.5971431732178 0.717455148696899
11.7585716247559 0.75731372833252
11.8299999237061 0.777243137359619
11.8671426773071 0.787207841873169
11.9700002670288 0.817101716995239
12.001428604126 0.827066421508789
12.0728569030762 0.846995711326599
12.1057138442993 0.856960296630859
12.1414289474487 0.866925001144409
12.2157144546509 0.886854290962219
12.2542858123779 0.896818995475769
12.294285774231 0.906783580780029
12.3814287185669 0.926712989807129
12.4314289093018 0.936677575111389
12.4842853546143 0.946642279624939
12.6071424484253 0.966571569442749
12.6942853927612 0.976536273956299
12.8228569030762 0.986500859260559
13.1042852401733 0.996465444564819
};
\addlegendentry{Height 8~m}
\addplot [thick, color1, dashed]
table {%
9.14142894744873 0
9.29714298248291 0.00996840000152588
9.3514289855957 0.0199368000030518
9.39428615570068 0.0299053192138672
9.5957145690918 0.0797474384307861
9.64000034332275 0.089715838432312
9.68142890930176 0.0996842384338379
9.7142858505249 0.109652638435364
9.75285720825195 0.11962103843689
9.8271427154541 0.139557957649231
9.85714244842529 0.149526357650757
9.88571453094482 0.159494757652283
9.91285705566406 0.169463157653809
9.94285678863525 0.179431676864624
9.96857166290283 0.18940007686615
9.99142837524414 0.199368476867676
10.0185718536377 0.209336876869202
10.0471429824829 0.219305276870728
10.0699996948242 0.229273796081543
10.0957145690918 0.239242196083069
10.1199998855591 0.249210596084595
10.1428575515747 0.259178996086121
10.16428565979 0.269147396087646
10.1871433258057 0.279115915298462
10.2114286422729 0.289084196090698
10.2299995422363 0.299052715301514
10.25 0.30902111530304
10.2771425247192 0.318989515304565
10.294285774231 0.328957915306091
10.3171424865723 0.338926315307617
10.3371429443359 0.348894834518433
10.3585710525513 0.358863234519958
10.3985710144043 0.37880003452301
10.4414281845093 0.398736953735352
10.458571434021 0.408705353736877
10.4785718917847 0.418673753738403
10.5214281082153 0.438610553741455
10.539999961853 0.448579072952271
10.5628566741943 0.458547472953796
10.582857131958 0.468515872955322
10.6071424484253 0.478484272956848
10.6300001144409 0.488452672958374
10.6542854309082 0.498421192169189
10.6742858886719 0.508389592170715
10.6971426010132 0.518357992172241
10.7214288711548 0.528326392173767
10.7442855834961 0.538294792175293
10.7957143783569 0.558231711387634
10.8171424865723 0.56820011138916
10.877142906189 0.588136911392212
10.9028568267822 0.598105430603027
10.9328575134277 0.608073830604553
10.9671430587769 0.618042230606079
10.997142791748 0.628010630607605
11.0257139205933 0.637979030609131
11.0614290237427 0.647947549819946
11.1000003814697 0.657915830612183
11.1300001144409 0.667884349822998
11.1700000762939 0.677852749824524
11.207142829895 0.68782114982605
11.2457141876221 0.697789669036865
11.2828569412231 0.707757949829102
11.375714302063 0.727694869041443
11.4214286804199 0.737663269042969
11.5200004577637 0.757600069046021
11.5642852783203 0.767568588256836
11.6171426773071 0.777536988258362
11.6742858886719 0.787505388259888
11.7842855453491 0.807442188262939
11.8500003814697 0.817410707473755
11.9128570556641 0.827379107475281
11.9714288711548 0.837347507476807
12.0342855453491 0.847316026687622
12.1057138442993 0.857284307479858
12.1800003051758 0.867252826690674
12.2399997711182 0.8772212266922
12.3199996948242 0.887189626693726
12.3971424102783 0.897158145904541
12.5600004196167 0.917094945907593
12.6614284515381 0.927063345909119
12.7399997711182 0.937031745910645
12.8228569030762 0.94700026512146
12.9214286804199 0.956968545913696
13.0299997329712 0.966937065124512
13.1342859268188 0.976905465126038
13.3185710906982 0.986873865127563
13.6785717010498 0.996842384338379
};
\addlegendentry{Height 11~m}
\addplot [thick, color2, dash pattern=on 1pt off 3pt on 3pt off 3pt]
table {%
8.43571472167969 0
8.6014289855957 0.010000467300415
8.61571407318115 0.0200009346008301
8.62285709381104 0.0300014019012451
8.62857151031494 0.0400019884109497
8.6371431350708 0.0600029230117798
8.6485710144043 0.100004911422729
8.64999961853027 0.110005378723145
8.65571403503418 0.130006313323975
8.65857124328613 0.150007247924805
8.66142845153809 0.16000771522522
8.66428565979004 0.180008769035339
8.66714286804199 0.190009236335754
8.68285751342773 0.300014495849609
8.68571472167969 0.310015082359314
8.69571399688721 0.380018472671509
8.69857120513916 0.390018939971924
8.70142841339111 0.410019874572754
8.70428562164307 0.420020341873169
8.70571422576904 0.430020809173584
8.71142864227295 0.450021862983704
8.71285724639893 0.460022330284119
8.71571445465088 0.470022797584534
8.71714305877686 0.480023264884949
8.7314281463623 0.530025720596313
8.73285675048828 0.540026187896729
8.74142837524414 0.570027589797974
8.74285697937012 0.580028176307678
8.74714279174805 0.590028643608093
8.77285671234131 0.680032968521118
8.77428531646729 0.690033435821533
8.78285694122314 0.720034956932068
8.78428554534912 0.730035424232483
8.78714275360107 0.740036010742188
8.791428565979 0.770037412643433
8.79428577423096 0.780037879943848
8.80142879486084 0.830040216445923
8.80428600311279 0.840040802955627
8.80714321136475 0.860041737556458
8.81285667419434 0.880042791366577
8.81714248657227 0.890043258666992
8.82285690307617 0.900043725967407
8.83142852783203 0.910044193267822
8.87428569793701 0.940045595169067
8.89285755157471 0.950046062469482
8.90999984741211 0.960046529769897
8.93000030517578 0.970047116279602
8.95142841339111 0.980047583580017
8.97999954223633 0.990048050880432
};
\addlegendentry{Height 24~m}
\end{axis}

\end{tikzpicture}
	\caption{CDFs of coherence bandwidth for different altitudes.}
	\label{fig:coh_band}
\end{figure}

Fig.~\ref{fig:rms_delay} shows the CDFs of the RMS delay spread for different heights calculated using Eq.~(\ref{eq:rms}). Overall, the ranges of RMS delay spread are $[259, 720]$~ns, $[207, 801]$~ns, $[348,552]$~ns for 8, 11, and 24~m, respectively. From statistical results, we can find that the $S_{\tau}$ has the largest standard deviation at 11~m, representing more obvious multipath effects. On the contrary, the large $\mu$ and small $\sigma$ of $S_{\tau}$ at 24~m demonstrate that MPCs have no large difference, and strong LOS propagation is not evident.  

Fig.~\ref{fig:coh_band} shows the CDFs of the coherence bandwidth for different altitudes. Overall, the ranges of coherence bandwidth are $[7.6, 13.9]$~MHz, $[9.1, 14.4]$~MHz, $[8.4, 9.5]$~MHz for 8, 11, and 24~m, respectively. Similar to $S_{\tau}$, $B_{coh}$ at 24~m is smaller than that of the other two heights. However, we found the SD and SA in the time domain are larger, which means the channel is characterized by low time variance but high frequency selectivity. Thus, time and frequency stationarities should be independently considered in the system design. %\zc{more insight?}

%The mean and standard deviation of the RMS delay spread and the coherence bandwidth can be found in TABLE~\ref{tab:stat}

% \begin{figure}
% 	\centering
%       \subfigure[Individual spatial correlation matrix for Element (3, 3) with pixels indicating the correlations between Element ($x$, $y$) and Element (3, 3), where we show the region (dashed line) for $\rho>0.85$ and the region (dashed and solid lines) for $\rho>0.80$.]{\input{img/dist_ac_90_individ_height_8m.tex}}
%         \label{fig:sc1}
%         \centering
%          \subfigure[Complete spatial correlation matrix for all 64 antennas. Where each pixel indicates the spatial correlation between two antenna elements.]{\includegraphics[width=1\linewidth]{img/dist_ac_90_height 8m.png}}
%         \label{fig:sc2}
%   % \end{subfigure}
% 	\caption{Spatial correlation of inter-elements when the UAV is at $90^{\circ}$ with respect to the BS and at a height of 8~m.}
% 	\label{fig:sc}
% \end{figure}

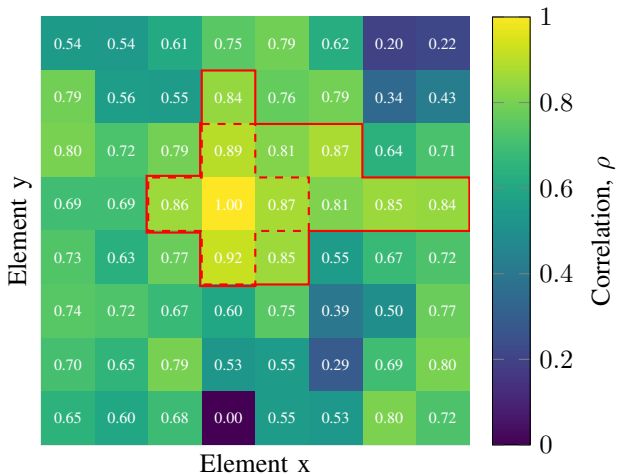
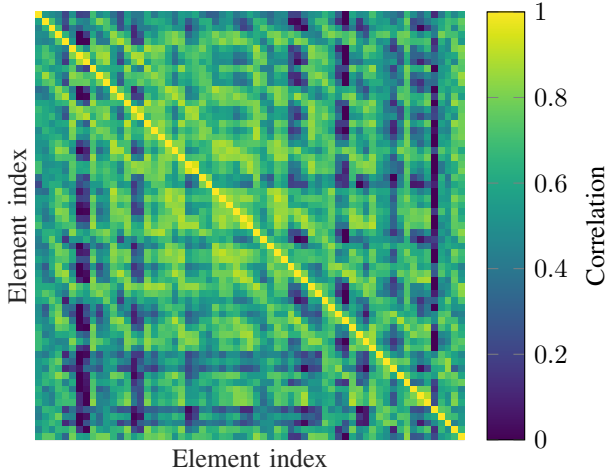
\begin{figure}
	\centering
    \begin{subfigure}[b]{0.9\linewidth}
        \centering
        % \begin{scaletikzpicturetowidth}{\linewidth}
            % This file was created with tikzplotlib v0.10.1.
\begin{tikzpicture}

\definecolor{darkslategray38}{RGB}{38,38,38}
%\definecolor{white!15!black}{RGB}{204,204,204}

\begin{axis}[
axis line style={white!80!black},
axis equal image,
colorbar,
colorbar style={ylabel={Correlation, $\rho$}},
colormap/viridis,
point meta max=1,
point meta min=0,
tick align=outside,
x grid style={white!80!black},
xlabel={Element x},
xmajorticks=false,
xmin=-0.5, xmax=7.5,
xtick style={color=black},
y dir=reverse,
y grid style={white!80!black},
ylabel={Element y},
ymajorticks=false,
ymin=-0.5, ymax=7.5,
ytick style={color=black}
]
\addplot graphics [includegraphics cmd=\pgfimage,xmin=-0.5, xmax=7.5, ymin=7.5, ymax=-0.5] {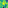};
\addplot [thick, red, dashed]
table {%
2.5 1.5
3.5 1.5
3.5 2.5
4.5 2.5
4.5 3.5
3.5 3.5
3.5 4.5
2.5 4.5
2.5 3.5
1.5 3.5
1.5 2.5
2.5 2.5
2.5 1.5
};
\addplot [thick, red]
table {%
2.5 1.5
2.5 0.5
3.5 0.5
3.5 1.5
5.5 1.5
5.5 2.5
7.5 2.5
7.5 3.5
4.5 3.5
4.5 4.5
3.5 4.5
3.5 4.53
2.47 4.53
2.47 3.53
1.47 3.53
1.47 2.47
2.47 2.47
2.47 1.47
};
\draw (axis cs:0,0) node[
  scale=0.6,
  text=white,
  rotate=0.0
]{0.54};
\draw (axis cs:1,0) node[
  scale=0.6,
  text=white,
  rotate=0.0
]{0.54};
\draw (axis cs:2,0) node[
  scale=0.6,
  text=white,
  rotate=0.0
]{0.61};
\draw (axis cs:3,0) node[
  scale=0.6,
  text=white,
  rotate=0.0
]{0.75};
\draw (axis cs:4,0) node[
  scale=0.6,
  text=white,
  rotate=0.0
]{0.79};
\draw (axis cs:5,0) node[
  scale=0.6,
  text=white,
  rotate=0.0
]{0.62};
\draw (axis cs:6,0) node[
  scale=0.6,
  text=white,
  rotate=0.0
]{0.20};
\draw (axis cs:7,0) node[
  scale=0.6,
  text=white,
  rotate=0.0
]{0.22};
\draw (axis cs:0,1) node[
  scale=0.6,
  text=white,
  rotate=0.0
]{0.79};
\draw (axis cs:1,1) node[
  scale=0.6,
  text=white,
  rotate=0.0
]{0.56};
\draw (axis cs:2,1) node[
  scale=0.6,
  text=white,
  rotate=0.0
]{0.55};
\draw (axis cs:3,1) node[
  scale=0.6,
  text=white,
  rotate=0.0
]{0.84};
\draw (axis cs:4,1) node[
  scale=0.6,
  text=white,
  rotate=0.0
]{0.76};
\draw (axis cs:5,1) node[
  scale=0.6,
  text=white,
  rotate=0.0
]{0.79};
\draw (axis cs:6,1) node[
  scale=0.6,
  text=white,
  rotate=0.0
]{0.34};
\draw (axis cs:7,1) node[
  scale=0.6,
  text=white,
  rotate=0.0
]{0.43};
\draw (axis cs:0,2) node[
  scale=0.6,
  text=white,
  rotate=0.0
]{0.80};
\draw (axis cs:1,2) node[
  scale=0.6,
  text=white,
  rotate=0.0
]{0.72};
\draw (axis cs:2,2) node[
  scale=0.6,
  text=white,
  rotate=0.0
]{0.79};
\draw (axis cs:3,2) node[
  scale=0.6,
  text=white,
  rotate=0.0
]{0.89};
\draw (axis cs:4,2) node[
  scale=0.6,
  text=white,
  rotate=0.0
]{0.81};
\draw (axis cs:5,2) node[
  scale=0.6,
  text=white,
  rotate=0.0
]{0.87};
\draw (axis cs:6,2) node[
  scale=0.6,
  text=white,
  rotate=0.0
]{0.64};
\draw (axis cs:7,2) node[
  scale=0.6,
  text=white,
  rotate=0.0
]{0.71};
\draw (axis cs:0,3) node[
  scale=0.6,
  text=white,
  rotate=0.0
]{0.69};
\draw (axis cs:1,3) node[
  scale=0.6,
  text=white,
  rotate=0.0
]{0.69};
\draw (axis cs:2,3) node[
  scale=0.6,
  text=white,
  rotate=0.0
]{0.86};
\draw (axis cs:3,3) node[
  scale=0.6,
  text=white,
  rotate=0.0
]{1.00};
\draw (axis cs:4,3) node[
  scale=0.6,
  text=white,
  rotate=0.0
]{0.87};
\draw (axis cs:5,3) node[
  scale=0.6,
  text=white,
  rotate=0.0
]{0.81};
\draw (axis cs:6,3) node[
  scale=0.6,
  text=white,
  rotate=0.0
]{0.85};
\draw (axis cs:7,3) node[
  scale=0.6,
  text=white,
  rotate=0.0
]{0.84};
\draw (axis cs:0,4) node[
  scale=0.6,
  text=white,
  rotate=0.0
]{0.73};
\draw (axis cs:1,4) node[
  scale=0.6,
  text=white,
  rotate=0.0
]{0.63};
\draw (axis cs:2,4) node[
  scale=0.6,
  text=white,
  rotate=0.0
]{0.77};
\draw (axis cs:3,4) node[
  scale=0.6,
  text=white,
  rotate=0.0
]{0.92};
\draw (axis cs:4,4) node[
  scale=0.6,
  text=white,
  rotate=0.0
]{0.85};
\draw (axis cs:5,4) node[
  scale=0.6,
  text=white,
  rotate=0.0
]{0.55};
\draw (axis cs:6,4) node[
  scale=0.6,
  text=white,
  rotate=0.0
]{0.67};
\draw (axis cs:7,4) node[
  scale=0.6,
  text=white,
  rotate=0.0
]{0.72};
\draw (axis cs:0,5) node[
  scale=0.6,
  text=white,
  rotate=0.0
]{0.74};
\draw (axis cs:1,5) node[
  scale=0.6,
  text=white,
  rotate=0.0
]{0.72};
\draw (axis cs:2,5) node[
  scale=0.6,
  text=white,
  rotate=0.0
]{0.67};
\draw (axis cs:3,5) node[
  scale=0.6,
  text=white,
  rotate=0.0
]{0.60};
\draw (axis cs:4,5) node[
  scale=0.6,
  text=white,
  rotate=0.0
]{0.75};
\draw (axis cs:5,5) node[
  scale=0.6,
  text=white,
  rotate=0.0
]{0.39};
\draw (axis cs:6,5) node[
  scale=0.6,
  text=white,
  rotate=0.0
]{0.50};
\draw (axis cs:7,5) node[
  scale=0.6,
  text=white,
  rotate=0.0
]{0.77};
\draw (axis cs:0,6) node[
  scale=0.6,
  text=white,
  rotate=0.0
]{0.70};
\draw (axis cs:1,6) node[
  scale=0.6,
  text=white,
  rotate=0.0
]{0.65};
\draw (axis cs:2,6) node[
  scale=0.6,
  text=white,
  rotate=0.0
]{0.79};
\draw (axis cs:3,6) node[
  scale=0.6,
  text=white,
  rotate=0.0
]{0.53};
\draw (axis cs:4,6) node[
  scale=0.6,
  text=white,
  rotate=0.0
]{0.55};
\draw (axis cs:5,6) node[
  scale=0.6,
  text=white,
  rotate=0.0
]{0.29};
\draw (axis cs:6,6) node[
  scale=0.6,
  text=white,
  rotate=0.0
]{0.69};
\draw (axis cs:7,6) node[
  scale=0.6,
  text=white,
  rotate=0.0
]{0.80};
\draw (axis cs:0,7) node[
  scale=0.6,
  text=white,
  rotate=0.0
]{0.65};
\draw (axis cs:1,7) node[
  scale=0.6,
  text=white,
  rotate=0.0
]{0.60};
\draw (axis cs:2,7) node[
  scale=0.6,
  text=white,
  rotate=0.0
]{0.68};
\draw (axis cs:3,7) node[
  scale=0.6,
  text=white,
  rotate=0.0
]{0.00};
\draw (axis cs:4,7) node[
  scale=0.6,
  text=white,
  rotate=0.0
]{0.55};
\draw (axis cs:5,7) node[
  scale=0.6,
  text=white,
  rotate=0.0
]{0.53};
\draw (axis cs:6,7) node[
  scale=0.6,
  text=white,
  rotate=0.0
]{0.80};
\draw (axis cs:7,7) node[
  scale=0.6,
  text=white,
  rotate=0.0
]{0.72};
\end{axis}

\end{tikzpicture}
        % \end{scaletikzpicturetowidth}
        \caption{Individual spatial correlation matrix for element (3, 3) with pixels indicating the correlations between element ($x$, $y$) and element (3, 3), where we show the region (dashed line) for $\rho>0.85$ and the region (dashed and solid lines) for $\rho>0.80$.}
        \label{fig:sc1}
    \end{subfigure}

    \begin{subfigure}[b]{0.9\linewidth}
        \centering
        % \begin{scaletikzpicturetowidth}{\linewidth}
            % This file was created with tikzplotlib v0.10.1.
\begin{tikzpicture}

\definecolor{darkslategray38}{RGB}{38,38,38}
%\definecolor{white!15!black}{RGB}{204,204,204}

\begin{axis}[
axis line style={white!80!black},
axis equal image,
colorbar,
colorbar style={ylabel={Correlation}},
colormap/viridis,
point meta max=1,
point meta min=0,
tick align=outside,
x grid style={white!80!black},
xlabel=\textcolor{darkslategray38}{Element index},
xmajorticks=false,
xmin=-0.5, xmax=62.5,
xtick style={color=black},
y dir=reverse,
y grid style={white!80!black},
ylabel=\textcolor{darkslategray38}{Element index},
ymajorticks=false,
ymin=-0.5, ymax=62.5,
ytick style={color=black}
]
\addplot graphics [includegraphics cmd=\pgfimage,xmin=-0.5, xmax=62.5, ymin=62.5, ymax=-0.5] {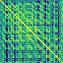};
\end{axis}

\end{tikzpicture}
        % \end{scaletikzpicturetowidth}
        \caption{Complete spatial correlation matrix for all 64 antennas. Where each pixel indicates the spatial correlation between two antenna elements.}
        \label{fig:sc2}
    \end{subfigure}

	\caption[Spatial correlation matrices of the antenna array.]{Spatial correlation matrices of the antenna array for a UAV at $90^{\circ}$ with respect to the base station and at a height of 8~m.}
	\label{fig:sc}
\end{figure}

\subsection{Spatial Stationarity}
The spatial stationarity is evaluated using the individual spatial correlation matrices and the complete spatial correlation matrix, which are shown for the location at 90~degrees at an altitude of $8~$m in Fig.~\ref{fig:sc}(a)  and Fig.~\ref{fig:sc}(b) respectively.

Fig.~\ref{fig:sc}(a) indicates the elements of the antenna array which have a correlation of $\rho > 0.85$ with a dashed line and the elements which have a correlation of $\rho > 0.80$ with a dashed or solid line. This gives a clear indication that the elements closest to the chosen antenna element have the largest correlation to it, which is expected. The threshold of $\rho$ is an empirically chosen value, further investigation is necessary to determine the stationarity region.

Fig.~\ref{fig:sc}(b) shows the correlation of each antenna element to all the other elements. This results in the correlation on the diagonal being one and we can see two offset diagonals with slightly higher correlation than the others, one 8 elements above and one 8 elements below the main diagonal. These correspond to the indices of the neighboring antenna elements above and below the main antenna element, so we can conclude that the behavior of the spatial correlation seen in Fig.~\ref{fig:sc}(a) is similar to all antenna elements.

\section{Conclusion and Future Work}
\label{sec:conclusion}
In this paper, we presented our contributions towards measurement system design, channel data collection and channel stationarity analysis of A2G MaMIMO channels. First, we introduced the first A2G MaMIMO channel measurement system using a lightweight USRP mounted on a drone and 64-antenna MaMIMO BS. Second, we used this measurement system to capture the previously unmeasured A2G MaMIMO channel. We conducted measurement campaigns to collect dense MaMIMO-UAV channel data at three different altitudes of UAV, 8~m, 11~m and 24~m.  
Third, we performed multidimensional channel stationarity analyses in three domains: temporal, frequency, and spatial stationarity. We first introduced a novel metric, stationary angle, which proves to be crucial for designing beamforming systems for large antenna arrays. This metric is supplementary to the traditional stationary distance to evaluate the temporal stationarity of the MaMIMO channel. In Section \ref{sec:sa} we show that the stationary angle is more meaningful when evaluating MaMIMO channels.
The results show that the SD ranges from 2.2-3.6~m, while the SA is 8.7-11.5$^\circ$, which provides a reference for physical layer design. 
We observed that the stationary angle at 24~m is 11.5~deg compared to just 8.7~deg at 11~m, indicating that the channel is more stable for higher-altitude UAVs.
For frequency stationarity, we found that the height at 24~m has a larger RMS delay spread, corresponding to a narrower coherence bandwidth. Moreover, the similar average scaling factor $\alpha$ ranging from 0.23 to 0.27 between these two parameters for three heights, indicates that this factor is environment-dependent. 
Finally, the spatial correlation was analyzed considering an individual element and the whole array. We found that by decreasing the correlation threshold from 0.85 to 0.8, the number of correlated elements in the vicinity of the main element increases from 4 to 11. This observation can be utilized for reducing channel estimation overheads in the spatial domain for MaMIMO systems. 

Although this paper studies comprehensive space-time-frequency channel behaviors at three typical heights of UAV (higher, lower, parallel), compared to the height of BS, more diverse trajectories should be considered in future work. Recently, we collected a more extensive data set featuring more complex trajectories (straight lines, circles) and different channel conditions (LOS and non-LOS). Following the analyses in this work, we will investigate A2G channels in more diverse trajectories using the recently collected data. This extensive data set is already open to the public \cite{colpaert23MaMIMO} for research use.

\balance

\bibliographystyle{IEEEtran}
\bibliography{new_references}
\end{document}